\documentclass[a4paper,11pt]{article}

\pdfoutput=1 

\usepackage{jheppub} 

\usepackage{slashed}
\usepackage{subcaption}
\usepackage{xcolor}
\usepackage{cleveref}
\usepackage[T1]{fontenc} 
\usepackage[utf8]{inputenc}
\usepackage{mathtools}

\DeclarePairedDelimiter\floor{\lfloor}{\rfloor}
\newcommand{\Lagr}{\mathcal{L}}

\newcommand{\cA}{\mathcal{A}}
\newcommand{\cM}{\mathcal{M}}

\title{\boldmath Tidal effects in quantum field theory}

\author{Kays Haddad}
\author{and Andreas Helset}
\affiliation{Niels Bohr International Academy and Discovery Center,
Niels Bohr Institute, \\ University of Copenhagen, Blegdamsvej 17,
DK-2100 Copenhagen, Denmark}
\emailAdd{kays.haddad@nbi.ku.dk}
\emailAdd{ahelset@nbi.ku.dk}

\abstract{We apply the Hilbert series to extend the gravitational action for a scalar field to 
a complete, non-redundant basis of higher-dimensional operators that is quadratic in the scalars and the Weyl tensor.
Such an extension of the action fully describes tidal effects arising from
operators involving two powers of the curvature.
As an application of this new action, we compute all spinless tidal effects
at the leading post-Minkowskian order
.
This computation is greatly simplified by appealing to the heavy limit,
where only a severely constrained set of operators can contribute classically at the one-loop level.
Finally, we use this amplitude to derive the $\mathcal{O}(G^{2})$
tidal corrections to the Hamiltonian and the scattering angle.
}

\subheader{\normalsize
        \vspace*{-3.7em}
        \begin{flushright}
                SAGEX-20-21-E
        \end{flushright}
}

\allowdisplaybreaks

\begin{document} 
\maketitle
\flushbottom

\section{Introduction}\label{sec:Intro}

There is a long history of relating scattering amplitudes
to conservative two-body classical observables.
Traditionally, such approaches have made extensive
use of the quantum action of gravity \cite{DeWitt},
and have been used most commonly to compute
non-relativistic classical and quantum corrections to
the interaction Hamiltonian \cite{Iwasaki,Donoghue:1993eb,Donoghue:1994dn,BjerrumBohr:2002kt,Holstein:2008sx}.
Other approaches still have utilized on-shell methods
to compute the amplitudes, before producing the
interaction Hamiltonian \cite{Neill:2013wsa,Bjerrum-Bohr:2013bxa,Vaidya:2014kza}.
Other than the interaction Hamiltonian,
refs.~\cite{BjerrumBohr:2002ks,Neill:2013wsa,Vaidya:2014kza}
also extracted information about the metric from the
two-to-two scattering amplitude.

Even compared to this illustrious record,
tremendous progress on this topic in a relatively short
time has been inspired by the detection of gravitational
waves (GWs) by the LIGO and Virgo collaborations \cite{LIGOGW}.
Developments in this time have by and large focused on
the post-Minkowskian (PM) expansion of amplitudes and
observables \cite{BERTOTTI1956,BERTOTTI1960169}.
This has required new tools for the conversion of PM
amplitudes to classical quantities such as the
interaction Hamiltonian \cite{Cheung:2018wkq,Cristofoli:2019neg,Bern:2020buy},
the linear and angular impulse and radiated momentum \cite{Kosower:2018adc,Maybee:2019jus},
the scattering angle \cite{Kalin:2019rwq,Bjerrum-Bohr:2019kec,Cristofoli:2020uzm},
and the metric \cite{Cristofoli:2020hnk}.
On the front of the amplitudes themselves, the current
state-of-the-art is the third post-Minkowskian (3PM)
amplitude for scalar-scalar scattering \cite{Zvi3PM,Bern:2019crd,Cheung:2020gyp}
(extended to include tidal effects in ref.~\cite{Cheung:2020sdj}).
The 3PM amplitude for massless scattering was also
computed in ref.~\cite{Bern:2020gjj}.
Moreover, amplitudes techniques have been used to
compute observables in modified theories of gravity
\cite{Brandhuber:2019qpg,Cristofoli:2019ewu}.

There has also been significant progress made on the
inclusion of spin effects.
The spin-1/2 $\times$ spin-1/2 amplitude was computed
up to the second post-Minkowskian order using
heavy particle effective theory (HPET) techniques in
ref.~\cite{Damgaard:2019lfh},
and was converted to a spinning Hamiltonian as part
of the spin-inclusive formalism of ref.~\cite{Bern:2020buy}.
An alternative approach to this amplitude involving the
leading singularity was presented in ref.~\cite{Guevara:2017csg}.
Making use of the massive on-shell variables of ref.~\cite{Arkani-Hamed:2017jhn},
several results including all orders in spin were achieved
in refs.~\cite{Guevara:2018wpp,Chung:2018kqs,Chung:2019duq,Guevara:2019fsj,Arkani-Hamed:2019ymq,Aoude:2020onz,Chung:2020rrz}.
Some of the notable results from these works include the interpretation of a
Kerr black hole as a minimally-coupled infinite spin particle,
the scattering angle at the second post-Minkowskian (2PM) order up to fourth order in spin,
an amplitudes interpretation of the Newman-Janis
complex deformation of Schwarzchild spacetime,
and the full 1PM spinning Hamiltonian.
Finally, ref.~\cite{Aoude:2020mlg} argued that the
scattering of minimally coupled spinning particles
minimizes the generated entanglement entropy.

Though a plethora of novel results have been achieved using
amplitudes-based approaches, the vast majority of results
directly applicable to GW templates have been derived
using general relativistic methods.
Of particular relevance to this paper are the computations
of tidal effects on the binary inspiral problem.
In this context, several tools have been applied to
the computations of these effects.
Two such tools are the post-Newtonian (PN) and PM approximations.
In the PN context, tidal moments were first
introduced in ref.~\cite{Damour:1992qi}.
Ref.~\cite{Damour:2009wj} incorporated tidal effects into 
the effective one-body (EOB) formalism \cite{Buonanno:1998gg}, and ref.~\cite{Bini:2012gu}
presented tidal contributions to the binding energy
within the EOB.
Most recently, tidal effects on the PM scattering angle
have been computed in refs.~\cite{Bini:2020flp,Kalin:2020mvi}

Up to this point, almost all amplitudes approaches
to the binary inspiral problem have ignored finite size and tidal
effects.
In fact, the recent work of ref.~\cite{Cheung:2020sdj} is the
first application of amplitudes methods to
the calculation of these effects.
By focusing on operators quadratic in the Weyl tensor,
they computed tidal contributions to spinless amplitudes
arising from the electric and magnetic quadrupoles,
up to the next-to-leading-PM order ($\mathcal{O}(G^{3})$).\footnote{Note that one-loop is the leading order where tidal effects can contribute to conservative dynamics.}
Converting their amplitudes to classical observables, they
found agreement with results derived from conventional
general relativistic methods \cite{Damour:2009wj,Bini:2012gu,Henry:2019xhg,Bini:2020flp,Kalin:2020mvi}.

In this paper, we expand on the work of ref.~\cite{Cheung:2020sdj}.
Through application of the Hilbert series (see e.g. \cite{Benvenuti:2006qr,Feng:2007ur,Jenkins:2009dy,Lehman:2015via,Lehman:2015coa,Henning:2015alf,Henning:2015daa,Henning:2017fpj,Ruhdorfer:2019qmk}),
we obtain a gravitational action comprising all operators
quadratic in the Weyl tensor and quadratic in a real scalar field.
This action is sufficient to fully describe all spinless tidal 
contributions to the amplitude at the leading-PM order ($\mathcal{O}(G^{2})$).
Since we are only interested in the classical portion of the amplitude,
we exploit the manifest $\hbar$ scaling of the heavy limit
of the action to isolate only classically
contributing operators \cite{Damgaard:2019lfh}.
This simplifies the computation, and we are 
able to straightforwardly produce the full classical tidal
integrand at the leading-PM order.
Integrating the integrand in principle requires knowledge of
the general even-rank triangle integral.
However, we are able circumvent this issue since we are simply 
interested in the leading-in-$\hbar$ portion of the integral that
is proportional to $S\equiv\pi^{2}/\sqrt{-q^{2}}$.
This allows us to find a form of the general even-rank triangle
integral that we have explicitly checked up to rank 10,
and that was proven in ref.~\cite{Bern:2020uwk} while this paper was in review.
Applying this results in the complete leading-PM tidal amplitude.
We indeed find the leading-PM contribution of ref.~\cite{Cheung:2020sdj}
as the leading contribution to our amplitude.
We then use our amplitude to derive all leading-PM tidal
corrections to the Hamiltonian and the scattering angle,
comparing to existing results along the way.

The layout of this paper is as follows:
We begin in \Cref{sec:TidalActions} by presenting the full
tidal actions for electromagnetism and gravity coupled to real
scalars at quadratic order in the field strength or the Weyl tensor respectively.
We include a brief primer on the Hilbert series in this section,
as it is the main tool in our construction.
With the tidal actions in hand, \Cref{sec:2PMTidal} focuses on the
computation of tidal contributions to the one-loop amplitudes.
The heavy limits of the tidal actions are also presented here.
We conclude in \Cref{sec:Conclusion}.

\section{Tidal actions}\label{sec:TidalActions}

This section is dedicated to the presentation of the
tidal actions up to quadratic order in the field
strengths or Weyl tensors respectively for QED or gravity coupled to a real scalar.
We achieve the complete forms of these actions through
application of the Hilbert series.
As such, we begin with a brief introduction to the
Hilbert series before presenting the results of the
series and corresponding tidal actions for QED and then gravity.
Technical details about the Hilbert series are postponed
to \Cref{sec:HilbertSeries}.

\subsection{Hilbert series for tidal effects}\label{sec:HSRev}

The Hilbert series uses character orthonormality to count group invariants. 
It is an important tool for constructing a basis of 
higher-dimensional operators, and has been applied to the effective-field-theory extension 
of the Standard Model in refs.~\cite{Lehman:2015via,Lehman:2015coa,Henning:2015daa,Henning:2015alf}, while ref.~\cite{Ruhdorfer:2019qmk} also included gravity.

The input for the Hilbert series is the field content and the fields' representations under
compact symmetries. The output is the number of invariant operators with a given field content 
and covariant derivatives. Redundancies coming from integration-by-parts relations and field redefinitions are taken into account.

We first want to construct operators with real scalar fields $\phi$ coupled to photons.
The Lorentz group $SO(1,3)$ is not a compact group, but we can use the Euclidean group 
$SO(4)\simeq SU(2)_L \times SU(2)_R$ to find the group invariants.
We then work with fields transforming in irreducible representations of $SU(2)_L$ and $SU(2)_R$ built from linear combinations
of the field strength $F_{\mu\nu}$ and the dual field strength
$\tilde F_{\mu\nu} = \frac{1}{2}\epsilon_{\mu\nu\rho\sigma}F^{\rho\sigma}$:
\begin{align}
F_{L/R}^{\mu\nu} \equiv \frac{1}{2}\left(F^{\mu\nu} \pm i\tilde F^{\mu\nu}\right).
\end{align}
The characters for $F_{L/R}^{\mu\nu}$ and $\phi$ are the input to the Hilbert series.

We restrict our attention to the operators with two real scalar fields, two field strengths, and an arbitrary 
number of covariant derivatives.
The output of the Hilbert series $\mathcal{H}_{d}^{F^2}$ for mass dimension $d=6+2n$ is
\begin{align}
	\label{eq:HilbertQEDrealScalar}
	\mathcal{H}_{6+2n}^{F^2} &= \floor*{\frac{n+2}{2}} \left( F_L^2 \phi^2 D^{2n} + F_R^2 \phi^2 D^{2n} \right)
	+ \floor*{\frac{n+1}{2}} F_L F_R \phi^2 D^{2n},
\end{align}
where $n\geq 0$ is an integer and $\floor*{x}$ is the floor function.

Now consider the Hilbert series for two real scalars coupled to gravity.
As explained in \Cref{sec:RedOps}, non-redundant operators quadratic
in the curvature can be written in terms of the Weyl
tensor $C^{\mu\nu\rho\sigma}$.
Thus we need only the group characters of
\begin{align}
C_{L/R}^{\mu\nu\rho\sigma} = \frac{1}{2}\left(C^{\mu\nu\rho\sigma} \pm i \tilde C^{\mu\nu\rho\sigma}\right),
\end{align}
where $\tilde C^{\mu\nu\rho\sigma} = \frac{1}{2}
\epsilon^{\mu\nu\alpha\beta}C_{\alpha\beta}^{\quad\rho\sigma}$ is the dual to the Weyl tensor.
The Hilbert series $\mathcal{H}_d^{C^2}$ for two real scalar fields, two Weyl tensors, and an arbitrary number of covariant derivatives is
\begin{align}
	\label{eq:HilbertGRrealScalar}
	\mathcal{H}_{6+2n}^{C^2} &= \floor*{\frac{n+2}{2}} \left( C_L^2 \phi^2 \nabla^{2n} + C_R^2 \phi^2 \nabla^{2n} \right)
	+ \floor*{\frac{n}{2}} C_L C_R \phi^2 \nabla^{2n},
\end{align}
for integer $n\geq 0$.

We use the output of the Hilbert series as a guide for constructing a basis of higher-dimensional
operators which capture all leading-PM tidal effects in electromagnetism and gravity.

\subsection{QED}\label{sec:QEDTidal}

The Lagrangian we are after couples a real scalar to
photons through operators quadratic in the field strength:
\begin{align}
	\label{eq:LagrQED}
	\mathcal{L}_{\rm QED} &= \frac{1}{2}\left( \partial_\mu \phi \right) \left( \partial^\mu \phi \right)
	- \frac{m^2}{2}\phi^2 +  \Delta \mathcal{L}_{\rm QED}^{\rm tidal}.
\end{align}
Here $\Delta \mathcal{L}_{\rm QED}^{\rm tidal}$ describes
the tidal interactions between two real scalars and two
field strength tensors.
We are interested in using the Hilbert series in \cref{eq:HilbertQEDrealScalar}
to construct this contribution at general mass dimension.

Ultimately, there is a freedom in the operator basis
we use (see \Cref{sec:RedOps}).
We choose a basis that is optimized for the computation
of classical amplitudes.
Such a basis does not include any structures of
the form $D^{\mu}\phi D_{\mu}\phi\mathcal{O}_{F^{2}}$.
These can be seen to mix with $\phi^{2}\mathcal{O}_{F^{2}}$
in the heavy limit, hence one could receive contributions
to classically-contributing heavy operators from an
infinite number of operators in the full action.
Furthermore, we will avoid derivative placements
that produce any structure that can be removed
by a field redefinition; see \Cref{sec:RedOps}
for a list of such structures.
Accounting for these criteria, we will build our
basis out of operators of the following form:
\begin{subequations}
\begin{align}
    \mathcal{O}_{LL,k}^{(n)}&=\left[D^{\mu_1 \dots \mu_k} \phi\right]
    \left[ D_{\nu_1 \dots \nu_k} \phi \right]
	\left[ D_{\mu_1 \dots \mu_k\alpha_1 \dots \alpha_{n-2k}} F_{L,\rho\sigma}\right]
	\left[ D^{\nu_1 \dots \nu_k\alpha_1 \dots \alpha_{n-2k}} F_{L}^{\rho\sigma}\right], \\
	\mathcal{O}_{RR,k}^{(n)}&=\left[D^{\mu_1 \dots \mu_k} \phi\right]
	\left[ D_{\nu_1 \dots \nu_k} \phi \right]
	\left[ D_{\mu_1 \dots \mu_k\alpha_1 \dots         \alpha_{n-2k}} F_{R,\rho\sigma}\right]
	\left[ D^{\nu_1 \dots \nu_k\alpha_1 \dots \alpha_{n-2k}} F_{R}^{\rho\sigma}\right], \\
	\mathcal{O}_{LR,k}^{(n+1)}&=\left[ D^{\rho\mu_1 \dots \mu_k} \phi \right]
	\left[ D_{\sigma\nu_1 \dots \nu_k} \phi \right]
	\left[ D_{\mu_1 \dots \mu_k\alpha_1 \dots \alpha_{n-2k}} F_{L,\rho\tau}\right]
	\left[ D^{\nu_1 \dots \nu_k\alpha_1 \dots \alpha_{n-2k}} F_{R}^{\sigma\tau}\right],
\end{align}
where $0\leq k\leq\lfloor n/2\rfloor$.
This range of $k$ produces the number of operators
dictated by the Hilbert series.
We have defined
$D_{\mu_1\dots \mu_n}\equiv D_{\mu_1}\dots D_{\mu_n}$.
\end{subequations}

We would like to construct our action out of
the fields $F^{\mu\nu}$ and $\tilde{F}^{\mu\nu}$.
To do so we simply replace $F_{L,R}^{\mu\nu}$ in terms
of the field strength and its dual.
After this replacement the operators above become
\begin{subequations}
\begin{align}
    \mathcal{O}_{LL,k}^{(n)}&=2\left[D^{\mu_1 \dots \mu_k} \phi\right]
    \left[ D_{\nu_1 \dots \nu_k} \phi \right]
	\left[ D_{\mu_1 \dots \mu_k\alpha_1 \dots \alpha_{n-2k}} F_{\rho\sigma}\right]
	\left[ D^{\nu_1 \dots \nu_k\alpha_1 \dots \alpha_{n-2k}} F^{\rho\sigma}\right]\notag \\
	&\quad+2i\left[D^{\mu_1 \dots \mu_k} \phi\right]
    \left[ D_{\nu_1 \dots \nu_k} \phi \right]
	\left[ D_{\mu_1 \dots \mu_k\alpha_1 \dots \alpha_{n-2k}} F_{\rho\sigma}\right]
	\left[ D^{\nu_1 \dots \nu_k\alpha_1 \dots \alpha_{n-2k}} \tilde{F}^{\rho\sigma}\right],\label{eq:LLOps} \\
	\mathcal{O}_{RR,k}^{(n)}&=2\left[D^{\mu_1 \dots \mu_k} \phi\right]
    \left[ D_{\nu_1 \dots \nu_k} \phi \right]
	\left[ D_{\mu_1 \dots \mu_k\alpha_1 \dots \alpha_{n-2k}} F_{\rho\sigma}\right]
	\left[ D^{\nu_1 \dots \nu_k\alpha_1 \dots \alpha_{n-2k}} F^{\rho\sigma}\right]\notag \\
	&\quad-2i\left[D^{\mu_1 \dots \mu_k} \phi\right]
    \left[ D_{\nu_1 \dots \nu_k} \phi \right]
	\left[ D_{\mu_1 \dots \mu_k\alpha_1 \dots \alpha_{n-2k}} F_{\rho\sigma}\right]
	\left[ D^{\nu_1 \dots \nu_k\alpha_1 \dots \alpha_{n-2k}} \tilde{F}^{\rho\sigma}\right],\label{eq:RROps} \\
	\mathcal{O}_{LR,k}^{(n+1)}&=2\left[ D^{\rho\mu_1 \dots \mu_k} \phi \right]
	\left[ D_{\sigma\nu_1 \dots \nu_k} \phi \right]
	\left[ D_{\mu_1 \dots \mu_k\alpha_1 \dots \alpha_{n-2k}} F_{\rho\tau}\right]
	\left[ D^{\nu_1 \dots \nu_k\alpha_1 \dots \alpha_{n-2k}} F^{\sigma\tau}\right]\notag \\
	&\quad-\frac{1}{2}\eta^{\rho\sigma}\left[ D^{\rho\mu_1 \dots \mu_k} \phi \right]
	\left[ D_{\sigma\nu_1 \dots \nu_k} \phi \right]
	\left[ D_{\mu_1 \dots \mu_k\alpha_1 \dots \alpha_{n-2k}} F_{\rho\tau}\right]
	\left[ D^{\nu_1 \dots \nu_k\alpha_1 \dots \alpha_{n-2k}} F^{\sigma\tau}\right]\label{eq:LROps}.
\end{align}
\end{subequations}
Both operators in \cref{eq:LLOps,eq:RROps} contain P-odd
terms.
We are not interested in such effects, so we ignore these
operators.
Also note that, by integrating by parts twice, the second 
term in \cref{eq:LROps} can be reexpressed in terms of
other operators already present and terms that can
be removed by field redefinitions, up to contributions
cubic in the field strength.

All-in-all, there are two generic structures out of
which we build the tidal action.
The tidal contribution to the action to all mass dimensions is thus
\begin{align}
	\label{eq:LagrQEDTidal}
	\Delta \mathcal{L}_{\rm QED}^{\rm tidal} &= 
	\sum_{n=0}^{\infty}\sum_{k=0}^{N} \left\{
		 a_{k}^{(n)} \left[D^{\mu_1 \dots \mu_k} \phi\right]
		\left[ D_{\nu_1 \dots \nu_k} \phi \right]
		\left[ D_{\mu_1 \dots \mu_k\alpha_1 \dots \alpha_{n-2k}} F_{\rho\sigma}\right]
		\left[ D^{\nu_1 \dots \nu_k\alpha_1 \dots \alpha_{n-2k}} F^{\rho\sigma}\right]
\right.\nonumber \\&
		+\left.  b_{k}^{(n+1)} 
		\left[ D^{\rho\mu_1 \dots \mu_k} \phi \right]
		\left[ D_{\sigma\nu_1 \dots \nu_k} \phi \right]
		\left[ D_{\mu_1 \dots \mu_k\alpha_1 \dots \alpha_{n-2k}} F_{\rho\tau}\right]
		\left[ D^{\nu_1 \dots \nu_k\alpha_1 \dots \alpha_{n-2k}} F^{\sigma\tau}\right]
	\right\}, 
\end{align}
where $N\equiv \floor*{n/2}$ and we have introduced the
Wilson coefficients $a_{k}^{(n)}$ and $b_{k}^{(n+1)}$.
Note that the covariant derivatives acting on the real scalars or field strenghts reduce
to partial derivatives.
One can easily incorporate P-odd operators into this
tidal action by including the same operators as in the first line in 
\cref{eq:LagrQEDTidal} where one of the field strenghts is replaced by a dual field strength.

\subsection{Gravity}\label{sec:GRTidal}

We repeat the procedure from the previous section, only
this time for a real scalar coupled to gravity.
The relevant action is
\begin{align}
	\label{eq:LagrGR}
	\sqrt{-g} \mathcal{L}_{\rm GR} &= \sqrt{-g} \left[
		\frac{g^{\mu\nu}}{2} \left( \partial_\mu \phi \right)
		\left( \partial_\nu \phi\right) - \frac{m^2}{2} \phi^2
		+  \Delta\mathcal{L}_{\rm GR}^{\rm tidal}
	\right].
\end{align}
We will find the form of the tidal contribution at
general mass dimension using the Hilbert series in \cref{eq:HilbertGRrealScalar}.

The optimal basis for our purposes satisfies the
same criteria as in the previous section.
As such, our basis comprises operators of the form
\begin{align}
    \mathcal{O}^{(n)}_{LL,k}&=\left[\nabla^{\mu_1 \dots \mu_k} \phi\right]
		\left[ \nabla_{\nu_1 \dots \nu_k} \phi \right]
		\left[ \nabla_{\mu_1 \dots \mu_k\alpha_1 \dots \alpha_{n-2k}} C_{L,\rho\sigma\alpha\beta}\right]
		\left[ \nabla^{\nu_1 \dots \nu_k\alpha_1 \dots \alpha_{n-2k}} C_{L}^{\rho\sigma\alpha\beta}\right], \\
    \mathcal{O}^{(n)}_{RR,k}&=\left[\nabla^{\mu_1 \dots \mu_k} \phi\right]
		\left[ \nabla_{\nu_1 \dots \nu_k} \phi \right]
		\left[ \nabla_{\mu_1 \dots \mu_k\alpha_1 \dots \alpha_{n-2k}} C_{R,\rho\sigma\alpha\beta}\right]
		\left[ \nabla^{\nu_1 \dots \nu_k\alpha_1 \dots \alpha_{n-2k}} C_{R}^{\rho\sigma\alpha\beta}\right], \\
    \mathcal{O}^{(n+2)}_{LR,k}&=\left[ \nabla^{\rho\sigma\mu_1 \dots \mu_k} \phi \right]
		\left[ \nabla_{\alpha\beta\nu_1 \dots \nu_k} \phi \right]
		\left[\nabla_{\mu_1 \dots \mu_k\alpha_1 \dots \alpha_{n-2k}} C_{L,\lambda\rho\tau\sigma}\right]
		\left[ \nabla^{\nu_1 \dots \nu_k\alpha_1 \dots \alpha_{n-2k}} C_{R}^{\lambda\alpha\tau\beta}\right].
\end{align}
We introduced the shorthand notation
$\nabla_{\mu_1\dots\mu_n}=\nabla_{\mu_1}\dots\nabla_{\mu_n}$.
In this case as well $k$ is in the range $0\leq k\leq N$.

These operators can be expressed in terms of the
Weyl tensor and its dual.
The exact same procedure as in the QED case,
along with covariant conservation of the Levi-Civita tensor \cite{Poplawski:2009fb}, results
in only two forms of operators comprising the basis,
modulo P-odd operators.
The tidal contribution to the action is thus
\begin{align}
	\label{eq:LagrGRTidal}
	&\Delta \mathcal{L}_{\rm GR}^{\rm tidal} = \\ &
	\sum_{n=0}^{\infty} \sum_{k=0}^{N} \left\{
		c_{k}^{(n)} \left[\nabla^{\mu_1 \dots \mu_k} \phi\right]
		\left[ \nabla_{\nu_1 \dots \nu_k} \phi \right]
		\left[ \nabla_{\mu_1 \dots \mu_k\alpha_1 \dots \alpha_{n-2k}} C_{\rho\sigma\alpha\beta}\right]
		\left[ \nabla^{\nu_1 \dots \nu_k\alpha_1 \dots \alpha_{n-2k}} C^{\rho\sigma\alpha\beta}\right]
\right.\nonumber \\&
		+\left.  d_{k}^{(n+2)} 
		\left[ \nabla^{\rho\sigma\mu_1 \dots \mu_k} \phi \right]
		\left[ \nabla_{\alpha\beta\nu_1 \dots \nu_k} \phi \right]
		\left[\nabla_{\mu_1 \dots \mu_k\alpha_1 \dots \alpha_{n-2k}} C_{\lambda\rho\tau\sigma}\right]
		\left[ \nabla^{\nu_1 \dots \nu_k\alpha_1 \dots \alpha_{n-2k}} C^{\lambda\alpha\tau\beta}\right]
	\right\}, \nonumber 
\end{align}
to all mass dimensions.
The coefficients $c_{k}^{(n)}$ and $d_{k}^{(n+2)}$
are the Wilson coefficients for the action.
Again, the P-odd operators which could be added to
the basis take the same form as the first line
in \cref{eq:LagrGRTidal} with one of the Weyl tensors replaced by a dual Weyl tensor.

\section{Tidal effects at the leading-PM order}\label{sec:2PMTidal}

The actions in \cref{eq:LagrQEDTidal,eq:LagrGRTidal}
describe all tidal effects that can arise from terms
quadratic in the electromagnetic field strength
and the curvature, respectively.
In fact, these actions are sufficient for describing all
tidal effects at the one-loop order, which corresponds
to the leading-PM order in the case of gravity.
We present in this section the full classical one-loop tidal
contributions to both electromagnetic and gravitational amplitudes.
Since we are exclusively interested in classical
contributions, we can take advantage of the heavy limits
of these actions to identify the only operators which
contribute classically at this loop order \cite{Damgaard:2019lfh}.
This results in significant simplifications to the
Feynman rules and the loop integrals involved.

\begin{figure}
    \centering
    \includegraphics[scale=1.2]{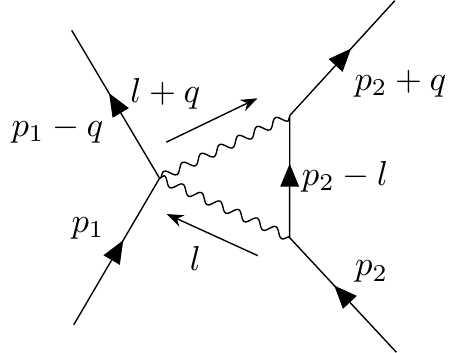}
    \caption{The only topology contributing classical tidal effects at one loop.
    The tidal effects of particle 1 are probed.
    The wavy lines represent either photons or gravitons.}
    \label{fig:Triangle}
\end{figure}

We follow the method in refs.~\cite{Guth:2014hsa,Namjoo:2017nia,Braaten:2018lmj,Damgaard:2019lfh}
to take the heavy limit of real scalars.
Namely, we apply the field redefinition
\begin{align}\label{eq:HeavyLimit}
    \phi\rightarrow\frac{1}{\sqrt{2m}}\left(e^{-imv\cdot x}\chi+e^{imv\cdot x}\chi^{*}\right),
\end{align}
and drop quickly oscillating terms.
Furthermore, by counting the powers of $\hbar$ associated
with each operator, and given that the triangle
diagram in \cref{fig:Triangle} is the only topology of
interest at the one-loop level, we only need the operators
at leading order in the $1/m$ expansion.

We present first the heavy limit of the electromagnetic
tidal action, as well as the full classical one-loop
tidal contribution to the electromagnetic amplitude, before moving on to the case of gravity.
We have normalized all amplitudes by multiplying by $4m_{1}m_{2}$.
This compensates for the normalization in \cref{eq:HeavyLimit}.\footnote{More precisely, this factor is the leading-in-$\hbar$ portion of the heavy scalar external states in momentum space \cite{Haddad:2020tvs}.}

\subsection{QED}\label{sec:QED2PM}

Beginning with the Lagrangian in \cref{eq:LagrQED},
we apply the field redefinition in \cref{eq:HeavyLimit} to obtain
\begin{subequations}\label{eq:LagrQEDHeavy}
\begin{align}
    \Lagr_{\mathrm{HQET}}&=\chi^{*}iv\cdot \partial\chi+\Delta\Lagr^{\mathrm{tidal}}_{\mathrm{HQET}}+\dots,
\end{align}
where
\begin{align}
	\label{eq:LagrQEDHeavyTidal}
    &\Delta\Lagr^{\mathrm{tidal}}_{\mathrm{HQET}}=\\&
    \sum_{n=0}^{\infty}\sum_{k=0}^{N}\left\{a_{k}^{(n)}m^{2k-1}
    [v^{\mu_{1}\dots \mu_{k}}\chi^*][v_{\nu_{1}\dots\nu_{k}}\chi]
    [D_{\mu_{1}\dots\mu_{k}\alpha_{1}\dots\alpha_{n-2k}}F_{\rho\sigma}]
    [D^{\nu_{1}\dots\nu_{k}\alpha_{1}\dots \alpha_{n-2k}}F^{\rho\sigma}]\right.\notag \\
    &
    \left.+b_{k}^{(n+1)}m^{2k+1}[v^{\rho\mu_{1}\dots\mu_{k}}\chi^{*}]
    [v_{\sigma\nu_{1}\dots\nu_{k}}\chi]
    [D_{\mu_{1}\dots\mu_{k}\alpha_{1}\dots\alpha_{n-2k}}F_{\rho\tau}]
    [D^{\nu_{1}\dots\nu_{k}\alpha_{1}\dots\alpha_{n-2k}}F^{\sigma\tau}]\right\}+\dots.\nonumber
\end{align}
\end{subequations}
We have defined $v_{\mu_1\dots\mu_n}=v_{\mu_1}\dots v_{\mu_n}$.
In these equations, dots represent operators
scaling with higher powers of $\hbar$.
We can ignore these operators as the computation of
classical effects at the one-loop level only requires
contributions from the leading-in-$\hbar$ operators.

At this point there is an apparent contradiction
in the tidal operators we have claimed to be leading in $\hbar$.
Increasing $n$ or decreasing $k$ at fixed $n$
increases the number of derivatives acting
on the photon field, thereby increasing powers of
$\hbar$ in the resulting contributions to amplitudes \cite{Kosower:2018adc}.
However, the derivative structure of the
subleading tidal terms in the worldline action
\cite{Bini:2012gu,Henry:2019xhg,Bini:2020flp}
suggests that we are right to keep these terms.
Therefore, much like in the case of spin effects where
the spin vector absorbs a power of $\hbar$ \cite{Maybee:2019jus}, we
propose that the tidal coefficients must scale with
$\hbar$ to absorb the factors from the operators.
The scalings that cancel those of the operators in \cref{eq:LagrQEDHeavy} are
\begin{subequations}
\begin{align}
    a_{k}^{(n)}&\sim\hbar^{-2n+2k-2}, \\
    b_{k}^{(n+1)}&\sim\hbar^{-2n+2k-2}.
\end{align}
\end{subequations}
As in the case of spin, this scaling is necessary to make contact between portions
of the amplitude and classical quantities.

We proceed now to use the heavy action to compute the
classical one-loop tidal amplitude.
We let particle $i$ have momentum $p_i = m_i v_i + k_i$,
where we have applied the usual heavy-particle decomposition of the 
momentum.
Note that we cannot generate three-point vertices
with a photon and two real scalars, so we take
particle 2 to be complex in this context, i.e. particle 2 
obeys the action given by eq.~(B.3) in ref.~\cite{Damgaard:2019lfh}.
The portion of the leading-PM amplitude involving only the
$k=0$ terms in \cref{eq:LagrQEDHeavy} is
\begin{align}
	\Delta\cA^{k=0}_{\rm 2} = -\frac{e^2}{\pi^2} S m_2 \sum_{n=0}^{\infty}(-1)^{n} \left(\frac{q^{2}}{2}\right)^{n+1}
	\left[ a_0^{(n)} + 
		\frac{m_1^2}{8} b_0^{(n+1)} (3\omega^2 + 1)
	\right].
\end{align}
We have defined $S\equiv\pi^{2}/\sqrt{-q^{2}}$ and $\omega\equiv v_{1}\cdot v_{2}$.
The notation $\Delta \mathcal{A}$ denotes an electromagnetic amplitude linear in the 
tidal coefficients in \cref{eq:LagrQEDHeavyTidal}.

Let's now extend this result to general $k$.
The unintegrated form of the amplitude is
\begin{align}
    \Delta\cA_{\rm 2}&=-8ie^{2}m_{2}\sum_{n=0}^{\infty}\sum_{k=0}^{N}(-1)^{n}m_{1}^{2k}\left(\frac{q^{2}}{2}\right)^{n-2k}\left[2\left(\frac{q^{2}}{2}\right)a_{k}^{(n)}v_{1\mu_{1}\dots\mu_{2k}}\mathcal{I}^{\mu_{1}\dots\mu_{2k}}_{\triangleleft}\right. \\
    &\qquad\qquad\qquad\qquad\quad
    \left.+m_{1}^{2}b_{k}^{(n+1)}\left(\omega^{2}\frac{q^{2}}{2}v_{\mu_{1}\dots\mu_{2k}}\mathcal{I}^{\mu_{1}\dots\mu_{2k}}_{\triangleleft}-v_{\mu_{1}\dots\mu_{2k+2}}\mathcal{I}^{\mu_{1}\dots\mu_{2k+2}}_{\triangleleft}\right)\right].\notag
\end{align}
To integrate the general $k$ amplitude, we need knowledge
of integrals of the form
\begin{align}
	\label{eq:genKint}
    v_{1\mu_{1}\dots\mu_{2k}}\mathcal{I}^{\mu_{1}\dots\mu_{2k}}_{\triangleleft}&=\int\frac{d^{4}l}{(2\pi)^{4}}\frac{(v_{1}\cdot l)^{2k}}{l^{2}(l+q)^{2}(-v_{2}\cdot l)}.
\end{align}
This task is simplified since we in fact only need the
portion of this integral proportional to the
non-analytic structure $S$,
and even then only the leading-in-$\hbar$ contribution to this portion.
We observe the following pattern for the portion of the
integral we are interested in:
\begin{align}\label{eq:GenkTriInt}
    v_{1\mu_{1}\dots\mu_{2k}}\mathcal{I}^{\mu_{1}\dots\mu_{2k}}_{\triangleleft}&=\frac{\left(\frac{1}{2}\right)_{k}}{4^{k}\left(1\right)_{k}}(\omega^{2}-1)^{k}q^{2k}\mathcal{I}_{\triangleleft}+\mathcal{O}(\hbar^{2k}),
\end{align}
where $(a)_{b}$ is the Pochhammer symbol and
\begin{align}
    \mathcal{I}_{\triangleleft}&=\int\frac{d^{4}l}{(2\pi)^{4}}\frac{1}{l^{2}(l+q)^{2}(-v_{2}\cdot l)}.
\end{align}
%
Note that, since the scalar triangle integral
scales as $\hbar^{-1}$, the leading term in \cref{eq:GenkTriInt}
scales as $\hbar^{2k-1}$.
We have explicitly checked \cref{eq:GenkTriInt} up to
$2k=10$ using the Passarino-Veltman reduction \cite{PASSARINO1979151}.
\Cref{eq:GenkTriInt} was proven in ref.~\cite{Bern:2020uwk} while this paper
was in review.

Armed with \cref{eq:GenkTriInt}, we compute a result
for general $n,k$;
%
\begin{align}
	\Delta\cA_{\rm 2} &=- \frac{e^{2}Sm_{2}}{\pi^{2}}\sum_{n=0}^{\infty}\sum_{k=0}^{N}(-1)^{n}\left(\frac{q^{2}}{2}\right)^{n-k+1}m_{1}^{2k}(\omega^{2}-1)^{k} \\
	&\quad\qquad\qquad\qquad
	\times\left\{a_{k}^{(n)}\frac{\left(\frac{1}{2}\right)_{k}}{2^{k}\left(1\right)_{k}}+\frac{1}{2^{k+1}}m_{1}^{2}b_{k}^{(n+1)}\left[\frac{\left(\frac{1}{2}\right)_{k}}{\left(1\right)_{k}}\omega^{2}-\frac{\left(\frac{1}{2}\right)_{k+1}}{2\left(1\right)_{k+1}}(\omega^{2}-1)\right]\right\}.\notag
\end{align}
%
We can reorganize the sums to make the dependence on $q^2$ more transparent:
\begin{align}
	\Delta\cA_{\rm 2} &= \sum_{i=0}^{\infty} \frac{e^2}{\pi^2} \left(-\frac{q^{2}}{2}\right)^{i+1}
	S m_2  f_i(\omega),
\end{align}
where
\begin{align}
	f_i(\omega) &\equiv
	\sum_{k=0}^{i} \frac{(\frac{1}{2})_k}{2^k(1)_k}(1-\omega^2)^k\left[ m_1^{2k} a_k^{(i+k)} 
	+\frac{1}{8(k+1)}m_1^{2k+2} b_k^{(i+k+1)}  \left[(2k+3)\omega^2 + (2k+1)\right]
	\right],
\end{align}
after some algebraic simplification of the Pochhammer symbols.

\subsection{Gravity}\label{sec:GR2PM}

We turn now to the leading-PM gravitational tidal amplitude.
Once again the first step is to find the action describing
a heavy scalar.
Beginning with the Lagrangian in \cref{eq:LagrGR}, we apply the field redefinition in \cref{eq:HeavyLimit}
to obtain
\begin{subequations}\label{eq:LagrGRHeavy}
\begin{align}
    \sqrt{-g}\Lagr_{\mathrm{HBET}}&=\sqrt{-g}\left[\frac{1}{2}m(g^{\mu\nu}v_{\mu}v_{\nu}-1)\chi^{*}\chi+\chi^{*}iv\cdot \partial\chi+\Delta\Lagr^{\mathrm{tidal}}_{\mathrm{HBET}}+\dots\right],
\end{align}
where
\begin{align}
	\label{eq:LagrGRHeavyTidal}
    &\Delta\Lagr^{\mathrm{tidal}}_{\mathrm{HBET}}= \\&
    \sum_{n=0}^{\infty}\sum_{k=0}^{N}\left\{c_{k}^{(n)}m^{2k-1}
    [v^{\mu_{1}\dots \mu_{k}}\chi^{*}][v_{\nu_{1}\dots\nu_{k}}\chi]
    [\nabla_{\mu_{1}\dots\mu_{k}\alpha_{1}\dots\alpha_{n-2k}}C_{\rho\sigma\alpha\beta}]
    [\nabla^{\nu_{1}\dots\nu_{k}\alpha_{1}\dots \alpha_{n-2k}}C^{\rho\sigma\alpha\beta}]\right.\notag \\
     &\quad
    \left.+d_{k}^{(n+2)}m^{2k+3}[v^{\rho\sigma\mu_{1}\dots\mu_{k}}\chi^{*}]
    [v_{\alpha\beta\nu_{1}\dots\nu_{k}}\chi]
    [\nabla_{\mu_{1}\dots\mu_{k}\alpha_{1}\dots\alpha_{n-2k}}C_{\lambda\rho\tau\sigma}]
    [\nabla^{\nu_{1}\dots\nu_{2}\alpha_{1}\dots\alpha_{n-2k}}C^{\lambda\alpha\tau\beta}]\right\}
	\nonumber \\ &\quad +\dots. \nonumber
\end{align}
\end{subequations}
In these equations, dots represent operators
scaling with higher powers of $\hbar$.
Note that we must keep the term $\chi^{*}iv\cdot\partial\chi$ in the action
even though it is subleading in $m$ since it is the
kinetic term for the heavy scalar.
Following the arguments in \cref{sec:QED2PM}, we propose the following $\hbar$-scaling of the 
gravitational tidal coefficients:
\begin{subequations}
\begin{align}
    c_{k}^{(n)}&\sim\hbar^{-2n+2k-4}, \\
    d_{k}^{(n+2)}&\sim\hbar^{-2n+2k-4}.
\end{align}
\end{subequations}

First we reproduce the leading-PM amplitude from ref.~\cite{Cheung:2020sdj}.
We need only the operators with $n=k=0$ for this task.
Thus the amplitude for the leading tidal effect is
\begin{align}
	\label{eq:2PMampGRleading}
	\Delta\cM^{n=k=0}_{\rm 2} &= G^2 q^4 S m_2^3 \left[ 
		16 c^{(0)}_0 + \frac{m_1^4}{8} d^{(2)}_0 (35 \omega^4 - 30 \omega^2 + 11) 
		\right].
\end{align}
This agrees with ref.~\cite{Cheung:2020sdj} with the identification $c^{(0)}_0\rightarrow \lambda/4$
and $d^{(2)}_0 \rightarrow \eta/(4m_1^4)$.
Here $\Delta \mathcal{M}$ is a gravitational amplitude linear in the tidal coefficients 
in \cref{eq:LagrGRHeavyTidal}.

We easily extend this result by including
terms at all orders in $n$ and with $k=0$:
\begin{align}
	\label{eq:2PMampGRalln}
	\Delta\cM^{k=0}_{\rm 2} &=4G^{2} S m_2^3  \sum_{n=0}^{\infty} (-1)^{n} \left(\frac{q^2}{2}\right)^{n+2} \left[ 
		16 c^{(n)}_0 + \frac{m_1^4}{8} d^{(n+2)}_0 (35 \omega^4 - 30 \omega^2 + 11) 
		\right].
\end{align}

The result for general $k$ depends on integrals of the form in \cref{eq:genKint}.
Specifically, the amplitude in terms of these integrals is
\begin{align}
    \Delta\cM_{\rm 2}&=512i\pi^{2}G^{2}m_{2}^{3}\sum_{n=0}^{\infty}\sum_{k=0}^{N}(-1)^{n}m_{1}^{2k}\left(\frac{q^{2}}{2}\right)^{n-2k+2}\left\{2c_{k}^{(n)}v_{\mu_{1}\dots\mu_{2k}}\mathcal{I}^{\mu_{1}\dots\mu_{2k}}_{\triangleleft}\right. \\
    &\left.+m_{1}^{4}d_{k}^{(n)}
    \left[
	q^{4}(1-2\omega^{2})^{2}v_{\mu_{1}\dots\mu_{2k}}\mathcal{I}^{\mu_{1}\dots\mu_{2k}}_{\triangleleft}
+4q^{2}(1-4\omega^{2})v_{\mu_{1}\dots\mu_{2k+2}}\mathcal{I}^{\mu_{1}\dots\mu_{2k+2}}_{\triangleleft}
	\right.\right.\nonumber \\ &\left.\left.\qquad\qquad
    +8v_{\mu_{1}\dots\mu_{2k+4}}\mathcal{I}^{\mu_{1}\dots\mu_{2k+4}}_{\triangleleft}
    \right]\right\}.\notag
\end{align}
We can integrate this using \cref{eq:GenkTriInt} to
obtain the result for all $n,k$:
\begin{align}
    &\Delta\cM_{\rm 2}=4G^{2}m_{2}^{3}S\sum_{n=0}^{\infty}\sum_{k=0}^{N}(-1)^{n}m_{1}^{2k}(\omega^{2}-1)^{k}\left(\frac{q^{2}}{2}\right)^{n-k+2}\left\{16c_{k}^{(n)}\frac{\left(\frac{1}{2}\right)_{k}}{2^{k}(1)_{k}}\right. \\
    &\left.+m_{1}^{4}d_{k}^{(n+2)}\left[
    (1-2\omega^{2})^{2}\frac{\left(\frac{1}{2}\right)_{k}}{2^{k-1}(1)_{k}}
    +(1-4\omega^{2})(\omega^2-1)\frac{\left(\frac{1}{2}\right)_{k+1}}{2^{k-1}(1)_{k+1}}
    +(\omega^{2}-1)^{2}\frac{\left(\frac{1}{2}\right)_{k+2}}{2^{k}(1)_{k+2}}
    \right]\right\}.\notag
\end{align}
%
A suggestive structure arises when $k\neq0$: 
each contribution is proportional to the factor
$(\omega^{2}-1)^{k}$. 
This factor is small in the PN limit, thus we can see already from the PM amplitude level that
the corresponding operators must be subleading in the PN limit, in 
agreement with the constructions in
refs.~\cite{Bini:2012gu,Henry:2019xhg,Bini:2020flp}.
In fact, this squares perfectly with principles from
classical gravitational effective field theories (EFTs).
Terms with $k\neq0$ involve derivatives of the Weyl tensor
of the form $v\cdot\nabla$. 
These reduce to time derivatives in the PN limit,
which are subleading compared to spatial derivatives
\cite{Goldberger:2004jt}.

Once again, we reorganize the sums in powers of the
transfer momentum.
The advantage of doing so is that contributions are
grouped by their significance to observables.
We find
\begin{subequations}\label{eq:GrAmpQExp}
\begin{align}
	\Delta\cM_{\rm 2} &= 4G^2 S  m_2^3 \sum_{i=0}^{\infty}  \left(-\frac{q^2}{2}\right)^{i+2}g_i(\omega),
\end{align}
where after some simplification
\newpage
\begin{align}
	g_i(\omega) &\equiv
	\sum_{k=0}^{i}\frac{(-1)^k(\frac{1}{2})_k}{2^k(1)_k}(\omega^2 - 1)^k\left[ 
		  16 m_1^{2k}c_{k}^{(i+k)} 
	\right.
	\\  &\qquad\qquad+\left.
		\frac{m_1^{2k+4}d^{(i+k+2)}_k}{ 4(k+2)(k+1)} \left[(2k+5)(2k+7)\omega^4 
	- 6(2k+5) \omega^2 
	+ (4k^2 + 12k + 11)\right] \right]. \nonumber
\end{align}
\end{subequations}
The amplitude is now presented in an optimal form for
conversion to the Hamiltonian or scattering angle.
We present these quantities in the next section, and defer comparison of this result with the literature until then.

In this section we have only computed the tidal contribution of particle 1 to the amplitude.
If one is interested in the tidal effects from both particles
at this order, one must simply symmetrize the results
here in the particle labels.

\subsection{Gravitational Hamiltonian and scattering angle}

We use now our leading-PM amplitude in \cref{eq:GrAmpQExp}
to compute the full leading-PM tidal corrections to the
Hamiltonian and the scattering angle.
Beginning with the Hamiltonian, there are two ways we may proceed.
The first is to match to the EFT of ref.~\cite{Cheung:2018wkq}, and the second is through the
Lippmann-Schwinger equation \cite{Cristofoli:2019ewu}.
As we are working to linear order in the tidal coefficients,
there will be no contributions from the Born iteration,
so we work here with the latter formulation.
The Hamiltonian as a function of the center-of-mass
momentum and the separation between the bodies is given by
\begin{align}
    H(\mathbf{p},\mathbf{r})&=\sum_{n=1,2}\sqrt{\mathbf{p}^{2}+m_{i}^{2}}+V(\mathbf{p},\mathbf{r})+\Delta V(\mathbf{p},\mathbf{r}).
\end{align}
Here $V(\mathbf{p},\mathbf{r})$ is the point particle potential and can be found
up to 3PM order in refs.~\cite{Zvi3PM,Bern:2019crd}.
$\Delta V(\mathbf{p},\mathbf{r})$ incorporates tidal
corrections.
At the order to which we have worked, these tidal corrections
are simply the Fourier transform of the leading-PM amplitude in the center-of-mass frame:
\begin{align}
    \Delta V(\mathbf{p},\mathbf{r})&=-\int\frac{d^{3}\mathbf{q}}{(2\pi)^{3}}e^{-i\mathbf{q}\cdot\mathbf{r}}\Delta\mathcal{M}_{2}(p,q).
\end{align}
In this frame the transfer momentum becomes $q^{\mu}=(0,\mathbf{q})$, so $q^{2}=-\mathbf{q}^{2}$.
Substituting now \cref{eq:GrAmpQExp} into this after 
incorporating the non-relativistic normalization $1/4E_{1}E_{2}$,
\begin{align}
    \Delta V(\mathbf{p},\mathbf{r})    &=-\frac{G^{2}m_{2}^{3}}{E^{2}\xi}\sum_{i=0}^{\infty}\frac{(-1)^{i}(2i+4)!}{2^{i+3}r^{2i+6}}g_{i}(\omega),
\end{align}
where $E\equiv E_{1}+E_{2}$ is the total energy in the center-of-mass frame
and $\xi\equiv E_{1}E_{2}/E^{2}$.
The $i=0$ term is in exact agreement with eq.~(10) of ref.~\cite{Cheung:2020sdj}.

With this in hand we can compute the scattering angle
using the method of ref.~\cite{Bjerrum-Bohr:2019kec}.
Note that $V_{\rm eff}$ in ref.~\cite{Bjerrum-Bohr:2019kec} is related
to the potential in position space by
$V_{\rm eff}=2E\xi\Delta V$.\footnote{We thank Andrea Cristofoli for pointing this out.}
Accounting for this, the scattering angle is
\begin{align}\label{eq:ScatAng}
    \Delta\chi&=\frac{G^{2}m_{2}^{3}}{E}\sum_{i=0}^{\infty}\frac{(-1)^{i}(2i+4)!(i+3)}{2^{i+2}p_{\infty}^{2}b^{2(i+3)}}\frac{\sqrt{\pi}\Gamma\left(i+\frac{7}{2}\right)}{\Gamma(i+4)}g_{i}(\omega),
\end{align}
where $b$ is the impact parameter and $p_{\infty}=|\mathbf{p}|$,
the magnitude of the center-of-mass three-momentum.
Evaluating this at $i=0$ and noting that $p_{\infty}b=J$, the angular momentum,
we find exact agreement with the $\mathcal{O}(J^{-6})$ portion of
eq.~(13) in ref.~\cite{Cheung:2020sdj}.
This also agrees with ref.~\cite{Bini:2020flp} upon converting to their notation and matching Wilson coefficients:
\begin{align}
p_{\infty}\rightarrow \frac{m_{1}m_{2}}{E}p_{\infty},&\quad
J\rightarrow Gm_{1}m_{2} j, \\
c_{0}^{(0)}\rightarrow-\frac{1}{12}m\sigma^{(2)},&\quad d_{0}^{(2)}\rightarrow\frac{1}{4m^{3}}\left(\mu^{(2)}+\frac{8}{3}\sigma^{(2)}\right).\label{eq:WilsCoMatchNoDeriv}
\end{align}
A similar notation conversion along with
the Wilson coefficient map in \cref{eq:WilsCoMatchNoDeriv} also produces agreement 
with the sum of eqs.~(5.5) and (5.6) of ref.~\cite{Kalin:2020mvi}.
We remark that the Wilson coefficient matching
in \cref{eq:WilsCoMatchNoDeriv} is equivalent
to the matching of ref.~\cite{Cheung:2020sdj}.
Moreover, this matching can be seen directly from the
level of the heavy action: it is the condition
that equates the $n=0,k=0$ portion of the heavy
tidal action \cref{eq:LagrGRHeavyTidal} with the
$l=2$ term of the classical worldline action in
ref.~\cite{Henry:2019xhg}, up to factors of $\chi^{*}\chi$.

To check the $i=1$ term we have repeated the calculation starting from the worldline action of
ref.~\cite{Bini:2020flp}, promoting each term to a quantum-field-theory operator, and multiplying by $\chi^{*}\chi$.
Doing so we find the following matching conditions on the Wilson coefficients of the
two operator bases:
%
\begin{subequations}\label{eq:J8Match}
\begin{align}
    c^{(1)}_{0}&\rightarrow-\frac{1}{32}m_{1}\sigma^{(3)}, \\
    c^{(2)}_{1}&\rightarrow\frac{1}{144m_{1}}\left(-\mu^{(3)}-12\sigma^{\prime(2)}+\frac{9}{2}\sigma^{(3)}\right), \\
    d^{(3)}_{0}&\rightarrow\frac{1}{12m_{1}^{3}}\left(\mu^{(3)}+3\sigma^{(3)}\right), \\
    d^{(4)}_{1}&\rightarrow\frac{1}{36m_{1}^{5}}\left(9\mu^{\prime(2)}-\mu^{(3)}+24\sigma^{\prime(2)}-3\sigma^{(3)}\right).
\end{align}
\end{subequations}
This mapping is also consistent with the form factors in eq.~(4.39) of
ref.~\cite{Bini:2020flp}, reproducing the same $\omega$ structure in $g_{i}(\omega)$
as in the form factors.\footnote{Note that $\omega$ in our notation is equivalent to $\gamma$ in the notation of ref.~\cite{Bini:2020flp}.}
Note, however, that this mapping is only appropriate up to overall constants when comparing
to the form factors, as we are comparing different quantities.

As a final check on the Wilson coefficient matching conditions,
we computed the factorizable portion of the tree-level $3\rightarrow3$ amplitude at linear order in the tidal coefficients.
Matching the amplitudes computed from both bases, we indeed find again the matching conditions in \cref{eq:WilsCoMatchNoDeriv,eq:J8Match}.

\section{Summary and outlook}\label{sec:Conclusion}

While the application of scattering amplitudes to the
binary point-particle inspiral problem has seen much
progress in recent years, the description of finite size 
and tidal effects is a novel and exciting development.
We have demonstrated the applicability of powerful EFT
tools to this problem.
Namely, through the Hilbert series we have been able to
write down an action which includes all possible
operators involving two real scalars and two Weyl tensors.
These operators represent the leading-PM tidal effects,
and the action they compose is sufficient to describe
all tidal contributions to the 2PM amplitude for
scalar-scalar scattering.

The computation of this amplitude was easily performed by
taking the heavy limit of the tidal action and isolating
only those operators with the correct $\hbar$ scaling to
contribute classically.
A subtlety arose in this identification of classically
contributing operators: operators with an increasing number
of derivatives acting on the Weyl tensors would have to
be considered classical.
This runs counter to the wisdom that more derivatives
produce more powers of $\hbar$.
To resolve this tension, we proposed 
that the Wilson coefficients of the action
must themselves scale with compensating powers of $\hbar$,
analogously to the absorption of $\hbar$ by the spin
vector.
We presented the unintegrated form of the leading-PM amplitude,
and integrated it using the form of the 
rank-$2k$ triangle integral in \cref{eq:GenkTriInt}.
We found agreement where our amplitude has overlap with existing results.
The amplitudes were then converted into a Hamiltonian and scattering angle, and once again
we found agreement with known results.

There are two obvious extensions to this work.
The first is the inclusion of tidal effects from operators with higher powers of the Weyl tensor.
An operator involving $n$ powers of the Weyl tensor contributes
to vertices with $n$ or more gravitons and two matter lines, and thus contributes to
conservative dynamics starting at the $n$PM order.
Second is the inclusion of spin effects.
This point is perhaps the more pressing of the two,
since tidal effects for objects of large enough spin may
also have implications for the Compton amplitude. 
The gravitational Compton amplitude acquires a 
spurious pole for matter with spin $s\geq2$ \cite{Arkani-Hamed:2017jhn},
an occurrence which is believed to derive from the 
necessarily composite nature of particles with large spin.
If this is true then it is natural to expect that the
inclusion of tidal effects may aid in remedying this
non-locality.
Both of these avenues can be pursued using the same Hilbert series methods we have employed here.
%
%
We leave these ideas for future research.

\acknowledgments

We thank Clifford Cheung, Andrea Cristofoli, Poul H. Damgaard, Mich\`ele Levi, Mikhail Solon, and
Matt von Hippel for related discussions.
We also thank Clifford Cheung and Mikhail Solon for comments on this manuscript.
This project has received funding from the European Union's Horizon 2020 
research and innovation programme under the Marie Sk\l{}odowska-Curie grant 
agreement No. 764850 "SAGEX".
The work of A.H. was supported in part by the Danish Research Foundation (DNRF91) and the Carlsberg 
Foundation.

\appendix

\section{Hilbert series}\label{sec:HilbertSeries}

Below we list the mathematical details we used in the construction 
of the Hilbert series for tidal effects.
For a detailed account of the Hilbert series, see e.g. refs.~\cite{Lehman:2015via,Lehman:2015coa,Henning:2015alf,Henning:2015daa,Henning:2017fpj,Jenkins:2009dy,Ruhdorfer:2019qmk}.

The Hilbert series $\mathcal{H}$ for a given field content $\phi$ is the contour integral of the plethystic exponential:\footnote{The modification term $\Delta \mathcal{H}$ will not be relevant for us as we consider
operators with mass dimension greater than 4.}
\begin{align}
	\mathcal{H} &= \int d\mu \frac{1}{P}{\rm PE}[\chi_{\phi}],
\end{align}
where the plethystic exponential (PE) generates all symmetric (antisymmetric) tensor products of the 
representations of the bosonic (fermionic) field content.
The factor $1/P$ removes a total derivative, where the momentum generating function $P$ is defined below in \cref{eq:MomGen}.
The plethystic exponential takes the form
\begin{align}
	{\rm PE}_\phi &= \exp\left[\sum_{r=0}^{\infty} z^{r+1}\frac{\phi^r}{r\mathcal{D}^{r\Delta_\phi}}\chi_\phi(x_1^r,\dots,x_k^r)\right],
\end{align}
where $\Delta_\phi$ is the mass dimension of $\phi$ and
$z=\pm 1$ when $\phi$ is a boson/fermion, respectively. Here $\chi_\phi$ is the character of the 
representation of $\phi$. 
When we consider several fields, we simply
multiply their plethystic exponentials.

We are using the Hilbert series to generate operators with neutral scalars, photons, and gravitons.
Thus we need their respective conformal representations:
\begin{align}
	\chi_\phi &= \chi_{[1,(0,0)]}(\mathcal{D};\alpha,\beta), \\
	\chi_{F_L} &= \chi_{[2,(1,0)]}(\mathcal{D};\alpha,\beta), \\
	\chi_{F_R} &= \chi_{[2,(0,1)]}(\mathcal{D};\alpha,\beta), \\
	\chi_{C_L} &= \chi_{[3,(2,0)]}(\mathcal{D};\alpha,\beta), \\
	\chi_{C_R} &= \chi_{[3,(0,2)]}(\mathcal{D};\alpha,\beta).
\end{align}
We could have included the characters for the $U(1)$ gauge group in electromagnetism, but, since both the scalars
and the photons are neutral, their characters would be trivial.

The characters for the unitary conformal representations of interest are \cite{Henning:2015alf,Henning:2017fpj,Barabanschikov:2005ri,Ruhdorfer:2019qmk}
\begin{align}
	\chi_{[1,(0,0)]}(\mathcal{D};\alpha,\beta) &= \mathcal{D} P(\mathcal{D};\alpha,\beta)(1-\mathcal{D}^2), \\
	\chi_{[3/2,(1/2,0)]}(\mathcal{D};\alpha,\beta) &= \mathcal{D}^{3/2} P(\mathcal{D};\alpha,\beta)\left[\chi_{(1/2,0)}(\alpha,\beta)-\mathcal{D}\chi_{(0,1/2)}(\alpha,\beta)\right], \\
	\chi_{[3/2,(0,1/2)]}(\mathcal{D};\alpha,\beta) &= \mathcal{D}^{3/2} P(\mathcal{D};\alpha,\beta)\left[\chi_{(0,1/2)}(\alpha,\beta)-\mathcal{D}\chi_{(1/2,0)}(\alpha,\beta)\right], \\
	\chi_{[2,(1,0)]}(\mathcal{D};\alpha,\beta) &= \mathcal{D}^2 P(\mathcal{D};\alpha,\beta)\left[\chi_{(1,0)}(\alpha,\beta)-\mathcal{D}\chi_{(1/2,1/2)}(\alpha,\beta)+\mathcal{D}^2\right], \\
	\chi_{[2,(0,1)]}(\mathcal{D};\alpha,\beta) &= \mathcal{D}^2 P(\mathcal{D};\alpha,\beta)\left[\chi_{(0,1)}(\alpha,\beta)-\mathcal{D}\chi_{(1/2,1/2)}(\alpha,\beta)+\mathcal{D}^2\right], \\
	\chi_{[3,(2,0)]}(\mathcal{D};\alpha,\beta) &= \mathcal{D}^3 P(\mathcal{D};\alpha,\beta)\left[\chi_{(2,0)}(\alpha,\beta)-\mathcal{D}\chi_{(3/2,1/2)}(\alpha,\beta)+\mathcal{D}^2\chi_{(1,0)}(\alpha,\beta)\right], \\
	\chi_{[3,(0,2)]}(\mathcal{D};\alpha,\beta) &= \mathcal{D}^3 P(\mathcal{D};\alpha,\beta)\left[\chi_{(0,2)}(\alpha,\beta)-\mathcal{D}\chi_{(1/2,3/2)}(\alpha,\beta)+\mathcal{D}^2\chi_{(0,1)}(\alpha,\beta)\right], 
\end{align}
where 
\begin{align}
	\label{eq:MomGen}
	P(\mathcal{D};\alpha,\beta) = \frac{1}{(1-\mathcal{D}\alpha\beta)(1-\mathcal{D}/(\alpha\beta))
	(1-\mathcal{D}\alpha/\beta)(1-\mathcal{D}\beta/\alpha)}
\end{align}
is the momentum generating function \cite{Henning:2015alf}. The characters of the Euclidean Lorentz
group are simply products of $SU(2)$ characters;
\begin{align}
	\chi_{(l_1,l_2)}(\alpha,\beta) = \chi_{l_1}^{SU(2)}(\alpha)\times \chi_{l_2}^{SU(2)}(\beta).
\end{align}
The $SU(2)$ characters we need are
\begin{align}
	\chi_{0}^{SU(2)}(\alpha) &= 1,  \\
	\chi_{1/2}^{SU(2)}(\alpha) &= \alpha + \frac{1}{\alpha}, \\ 
	\chi_{1}^{SU(2)}(\alpha) &= \alpha^2 +  1 + \frac{1}{\alpha^2},  \\ 
	\chi_{3/2}^{SU(2)}(\alpha) &= \alpha^3 + \alpha + \frac{1}{\alpha} + \frac{1}{\alpha^3},  \\
	\chi_{2}^{SU(2)}(\alpha) &= \alpha^4 + \alpha^2 + 1 + \frac{1}{\alpha^2} + \frac{1}{\alpha^4}. 
\end{align}
Finally, the Haar measure for the Euclidean Lorentz group $SO(4)\simeq SU(2)_L \times SU(2)_R$ is
\begin{align}
	\int d\mu_{\rm Lorentz} =  \left(\frac{1}{2\pi i}\right)^2 
	\oint_{|\alpha|=1} \frac{d\alpha}{2\alpha} (1-\alpha^2)\left(1-\frac{1}{\alpha^2}\right)
	\oint_{|\beta|=1} \frac{d\beta}{2\beta} (1-\beta^2)\left(1-\frac{1}{\beta^2}\right).
\end{align}

\section{Redundant operators}\label{sec:RedOps}

The operator basis for leading-PM tidal effects in \cref{eq:LagrGRTidal} is 
a complete, non-redundant basis for all operators involving
two scalars and two Weyl tensors. However, the explicit form of
the operator basis involves some choices, originating from two 
types of redundancies: field redefinitions and integration-by-parts relations.

First we consider redundancies from field redefinitions.
The free equation of motion (EOM) for the scalar field is
\begin{align}
	\partial^2 \phi + m^2 \phi = 0.
\end{align}
A composite operator which contains the factor $\partial^2 \phi$ can be removed from the operator basis
by an appropriate choice of field redefinition which exchanges it for the operator $m^{2}\phi$.
When constructing operators where partial derivatives
are acting on the scalar fields, we need only consider symmetric, traceless combinations of the derivatives.\footnote{We could also replace the partial derivatives with 
	covariant derivatives. Note that commutators of covariant derivatives are related to field strengths or
curvature; $[D_\mu, D_\nu] \sim F_{\mu\nu}$ or $[\nabla_\mu, \nabla_\nu] \sim R$.}

Similarly, the free EOM for the gauge field is
\begin{align}
	\partial_\mu F^{\mu\nu} = 0.
\end{align}
Also, we find that 
\begin{align}
	\partial^2 F_{\mu\nu} = 0
\end{align}
by using the Bianchi identity $\partial_{[\alpha}F_{\mu\nu]} = 0 $.
Again, we only need symmetric, traceless combinations of derivatives acting on the field strengths, where none
of the derivatives are contracted with that field strength.

For gravity, we have Einstein's equation in vacuum:
\begin{align}
	R_{\mu\nu} = 0,
\end{align}
where $R_{\mu\nu}$ is the Ricci tensor.
This means that we don't include the Ricci tensor nor the Ricci scalar in the operator basis as they can be 
removed by an appropriate redefinition of the metric tensor.\footnote{From an amplitude perspective, ref.~\cite{Huber:2019ugz} showed that the modification of
the Einstein-Hilbert action by the addition of $R^{2}$ and $R^{\mu\nu}R_{\mu\nu}$ terms does not affect the amplitude.
Ref.~\cite{Henry:2019xhg} found an explicit field redefinition of the graviton field that removes traces of
the Riemann curvature from the tidal worldline action,
including in the presence of matter.}

Since the Weyl tensor is the traceless part of the Riemann tensor,
\begin{align}
	C_{\mu\nu\rho\sigma} = R_{\mu\nu\rho\sigma} - \left(g_{\mu[\rho}R_{\sigma]\nu} - g_{\nu[\rho}R_{\sigma]\mu}\right) + \frac{1}{3}g_{\mu[\rho}g_{\sigma]\nu}R,
\end{align}
where $A_{[\mu\nu]}=\frac{1}{2}(A_{\mu\nu}-A_{\nu\mu})$ for any tensor $A$, we can freely work with either 
the Riemann tensor or the Weyl tensor. For our purposes, it is most convenient to work with the Weyl tensor 
because it transforms in an irreducible representation of the Euclidean Lorentz group; see \Cref{sec:HilbertSeries}.

In vacuum, we find that 
\begin{align}
	\nabla^{\mu} C_{\mu\nu\rho\sigma} &= 0, \\ 
	\nabla^2 C_{\mu\nu\rho\sigma} &= \mathcal{O}(C^2),
\end{align}
up to terms with Ricci tensors or Ricci scalars. Thus we need only keep symmetric, traceless combinations of covariant derivatives acting on the Weyl tensors.

Next we will illustrate the redundancies coming from integration-by-parts relations 
by looking at some possible dimension-8 operators:
\begin{align}
	\mathcal{O}_1 &= \phi \phi \left[\nabla_\mu C_{\rho\sigma\alpha\beta}\right]\left[\nabla^{\mu} C^{\rho\sigma\alpha\beta}\right], \\
	\mathcal{O}_2 &= \left[ \nabla_\mu \phi\right]\left[\nabla^\mu \phi\right]  C_{\rho\sigma\alpha\beta} C^{\rho\sigma\alpha\beta}, \\
	\mathcal{O}_3 &= \phi\left[\nabla^\mu \phi\right]  \left[ \nabla_\mu C_{\rho\sigma\alpha\beta} \right]C^{\rho\sigma\alpha\beta}.
\end{align}
Here, $\mathcal{O}_1$ corresponds to the operator with coefficient $c_{0}^{(1)}$ in \cref{eq:LagrGRTidal}, while 
$\mathcal{O}_2$ and $\mathcal{O}_3$ are absent from \cref{eq:LagrGRTidal}. The Hilbert series in \cref{eq:HilbertGRrealScalar}
informs us that there should be only one P-even operator at this mass dimension, but it does not tell us
which one we should choose.

In fact, these operators are related through integration-by-parts relations,
\begin{align}
\mathcal{O}_1 &= \mathcal{O}_2 + \mathcal{\rm EOM} + \mathcal{O}(C^3), \\
\mathcal{O}_{1} &= -\mathcal{O}_{3},
\end{align}
up to a total derivative, operators proportional to the leading-order EOM, 
and operators with more than two Weyl tensors.
We discard the total derivative due to momentum conservation, and the EOM operators can be removed through
a field redefinition. In fact, the Hilbert series have implicitly removed, whenever possible, operators with
more derivatives in place of operators with fewer derivatives, i.e. using the EOM.

When we have more than one operator at a given mass dimension, we must carefully include independent operators
which cannot be related through integration-by-parts relations or field redefinitions.
A systematic way of taking into account integration-by-parts relations is detailed in refs.~\cite{Lehman:2015coa,Hays:2018zze}.
We enumerate all the the ways of partitioning the derivatives (ignoring integration-by-parts relations), which we call $\{x_i\}$.
Then we enumerate all gauge-invariant operators with
one fewer covariant derivative which transform as a Lorentz four-vector, $\{y_i\}$.
We can then apply a total derivative to the $y_i$'s, which will generate a relation among the $x_i$'s.
The number of independent constraints coming from this procedure is given by the rank of the matrix
of constraint equations.

Let's illustrate the procedure for the dimension-8 operators. We assign the $x_i = \mathcal{O}_i$ for $i=1,2,3$.
For the operators with one covariant derivative, we can have the covariant derivative act on a scalar or on a Weyl tensor;
\begin{align}
	y_{1,\mu} &= \phi \left[\nabla_\mu \phi \right] C_{\rho\sigma\alpha\beta} C^{\rho\sigma\alpha\beta}, \\
	y_{2,\mu} &= \phi \phi \left[\nabla_\mu C_{\rho\sigma\alpha\beta}\right] C^{\rho\sigma\alpha\beta}. 
\end{align}
Now we apply the total derivative on the $y_i$'s:
\begin{align}
	\nabla^{\mu}y_{1,\mu} &= x_1 + 2 x_2 = 0,  \\
	\nabla^{\mu}y_{2,\mu} &= 2 x_2 + x_3 = 0.  
\end{align}
Note that we have dropped operators with $D^2 \phi$ or $D^2 C$ because they can either be removed by
field redefinitions or produce operators with more than two Weyl tensors.
We can write the equations in matrix form,
\begin{align}
	M.x \equiv  \begin{pmatrix}
	1 & 2 & 0 \\
	0 & 2 & 1 
	\end{pmatrix} 
	\begin{pmatrix}
		x_1 \\ x_2 \\ x_3
	\end{pmatrix} = 0.
\end{align}
The number of independent operators is $3-{\rm rank}(M)=3-2=1$.

We have applied this method to ensure the operators in our basis are independent up to mass dimension 14.
For the higher mass dimensions, we used the on-shell methods discussed in \Cref{sec:HSOnShell}.

We have illustrated the freedom in choosing an operator basis coming from integration-by-parts relations.
However, certain operator bases are better suited for calculations in the heavy limit.
For example, the heavy limit of $\mathcal{O}_2$ feeds down to the dimension-6 operator $\phi\phi C_{\rho\sigma\alpha\beta}C^{\rho\sigma\alpha\beta}$,
so this operator doesn't contribute new information with regards to the classical portion of the amplitude.
In fact, we would also need to include subleading-in-$\hbar$ corrections from $\mathcal{O}_2$ to reproduce the correct
subleading tidal effects.

The operator basis in \cref{eq:LagrGRTidal} is chosen to optimally produce all leading-PM tidal effects in the classical limit.

\section{Operator basis from an on-shell perspective}\label{sec:HSOnShell}

A different approach to constructing the operator basis is to first look at the corresponding 
on-shell amplitudes. Following the discussion in ref.~\cite{Shadmi:2018xan}, we consider the 
non-factorizable part of the two-scalar-two-photon amplitude $\mathcal{A}(\phi \phi;\gamma \gamma)$.
We label the momenta for the photons by $p_1$ and $p_2$, and the momenta of the massive scalars by
$p_3$ and $p_4$.
For the helicity assignments $\gamma^{+}(p_{1})\gamma^{+}(p_{2})$
and $\gamma^{-}(p_{1})\gamma^{+}(p_{2})$,
the structures carrying the correct little group weights are
\begin{align}
	[12] \qquad \textrm{and} \qquad \langle 1|(p_3 - p_4)| 2],
\end{align}
respectively.
The amplitudes for the other helicity assignments can be constructed from the same building blocks after exchanging angle and square brackets.
The non-factorizable part of the two amplitudes with positive helicity for $p_2$ are 
\begin{align}
	\mathcal{A}(\phi\phi; \gamma^{+}(p_1),\gamma^{+}(p_2)) &= [12]^2 a(s_{12},s_{13},s_{14}),	 \\
	\mathcal{A}(\phi\phi; \gamma^{-}(p_1),\gamma^{+}(p_2)) &= \langle 1| (p_3 - p_4)|2]^2 	b(s_{12},s_{13},s_{14}),
\end{align}
where $a(s_{12},s_{13},s_{14})$ and $b(s_{12},s_{13},s_{14})$ are polynomials of the 
Mandelstam variables $s_{ij}=(p_i+p_j)^2$. 
Taking into account the relation $s_{12}+s_{13}+s_{14}=2m^2$, and keeping the symmetry $3\leftrightarrow 4$,
the polynomials take the form
\begin{align}
	a(s_{12},s_{13},s_{14}) &= \sum_{i=0}^{\infty} \sum_{j=0}^{\infty} \frac{a_{i,j}}{\Lambda^{2i+4j+2}} s_{12}^i (s_{13} s_{14})^j, \\
	b(s_{12},s_{13},s_{14}) &= \sum_{i=0}^{\infty} \sum_{j=0}^{\infty} \frac{b_{i,j}}{\Lambda^{2i+4j+4}} s_{12}^i (s_{13} s_{14})^j,
\end{align}
where $\Lambda$ is some unfixed dimensionful scale and $a_{i,j}$ and $b_{i,j}$ are dimensionless 
coefficients.

By comparing the non-factorizable part of the on-shell amplitudes with the output of the Hilbert series
in \cref{eq:HilbertQEDrealScalar},
one can find a correspondence between the Wilson coefficients of the action
and the coefficients $a_{i,j},$ $b_{i,j}$.
This helps us in inferring the higher-dimensional operators, since we can now construct operators 
which have the field content given by the Hilbert series and which reduce to the amplitudes
when imposing on-shell conditions.

Similarly, we can compare the on-shell amplitudes for two scalars and two gravitons,
\begin{align}
	\cM(\phi\phi;g^{2+}(p_1),g^{2+}(p_2)) &= [12]^4 c(s_{12},s_{13},s_{14}), \\
	\cM(\phi\phi;g^{2-}(p_1),g^{2+}(p_2)) &= \langle 1|(p_3 - p_4)|2]^4 d(s_{12},s_{13},s_{14}),
\end{align}
with the output of the Hilbert series in \cref{eq:HilbertGRrealScalar}.
The same arguments apply to the polynomials $c$ and $d$ as $a$ and $b$, so they
become
\begin{align}
	c(s_{12},s_{13},s_{14}) &= \sum_{i=0}^{\infty} \sum_{j=0}^{\infty} \frac{c_{i,j}}{\Lambda^{2i+4j+4}} s_{12}^i (s_{13} s_{14})^j, \\
	d(s_{12},s_{13},s_{14}) &= \sum_{i=0}^{\infty} \sum_{j=0}^{\infty} \frac{d_{i,j}}{\Lambda^{2i+4j+8}} s_{12}^i (s_{13} s_{14})^j,
\end{align}
with dimensionless coefficients $c_{i,j}$ and $d_{i,j}$. 
We see a similar correspondence between the on-shell amplitudes and the effective operators
as in the QED case.

\bibliographystyle{JHEP}
\bibliography{TidalEffects}

\providecommand{\href}[2]{#2}\begingroup\raggedright\begin{thebibliography}{10}

\bibitem{DeWitt}
B.~S. DeWitt, \emph{Quantum theory of gravity. ii. the manifestly covariant
  theory}, \href{https://doi.org/10.1103/PhysRev.162.1195}{\emph{Phys. Rev.}
  {\bfseries 162} (1967) 1195}.

\bibitem{Iwasaki}
Y.~Iwasaki, \emph{{Quantum Theory of Gravitation vs. Classical Theory*):
  Fourth-Order Potential}},
  \href{https://doi.org/10.1143/PTP.46.1587}{\emph{Progress of Theoretical
  Physics} {\bfseries 46} (1971) 1587}
  [\href{https://arxiv.org/abs/https://academic.oup.com/ptp/article-pdf/46/5/1587/5271183/46-5-1587.pdf}{{\ttfamily
  https://academic.oup.com/ptp/article-pdf/46/5/1587/5271183/46-5-1587.pdf}}].

\bibitem{Donoghue:1993eb}
J.~F. Donoghue, \emph{{Leading quantum correction to the Newtonian potential}},
  \href{https://doi.org/10.1103/PhysRevLett.72.2996}{\emph{Phys. Rev. Lett.}
  {\bfseries 72} (1994) 2996}
  [\href{https://arxiv.org/abs/gr-qc/9310024}{{\ttfamily gr-qc/9310024}}].

\bibitem{Donoghue:1994dn}
J.~F. Donoghue, \emph{{General relativity as an effective field theory: The
  leading quantum corrections}},
  \href{https://doi.org/10.1103/PhysRevD.50.3874}{\emph{Phys. Rev.} {\bfseries
  D50} (1994) 3874} [\href{https://arxiv.org/abs/gr-qc/9405057}{{\ttfamily
  gr-qc/9405057}}].

\bibitem{BjerrumBohr:2002kt}
N.~E.~J. Bjerrum-Bohr, J.~F. Donoghue and B.~R. Holstein, \emph{{Quantum
  gravitational corrections to the nonrelativistic scattering potential of two
  masses}}, \href{https://doi.org/10.1103/PhysRevD.71.069903,
  10.1103/PhysRevD.67.084033}{\emph{Phys. Rev.} {\bfseries D67} (2003) 084033}
  [\href{https://arxiv.org/abs/hep-th/0211072}{{\ttfamily hep-th/0211072}}].

\bibitem{Holstein:2008sx}
B.~R. Holstein and A.~Ross, \emph{{Spin Effects in Long Range Gravitational
  Scattering}},  \href{https://arxiv.org/abs/0802.0716}{{\ttfamily 0802.0716}}.

\bibitem{Neill:2013wsa}
D.~Neill and I.~Z. Rothstein, \emph{{Classical Space-Times from the S Matrix}},
  \href{https://doi.org/10.1016/j.nuclphysb.2013.09.007}{\emph{Nucl. Phys. B}
  {\bfseries 877} (2013) 177}
  [\href{https://arxiv.org/abs/1304.7263}{{\ttfamily 1304.7263}}].

\bibitem{Bjerrum-Bohr:2013bxa}
N.~E.~J. Bjerrum-Bohr, J.~F. Donoghue and P.~Vanhove, \emph{{On-shell
  Techniques and Universal Results in Quantum Gravity}},
  \href{https://doi.org/10.1007/JHEP02(2014)111}{\emph{JHEP} {\bfseries 02}
  (2014) 111} [\href{https://arxiv.org/abs/1309.0804}{{\ttfamily 1309.0804}}].

\bibitem{Vaidya:2014kza}
V.~Vaidya, \emph{{Gravitational spin Hamiltonians from the S matrix}},
  \href{https://doi.org/10.1103/PhysRevD.91.024017}{\emph{Phys. Rev. D}
  {\bfseries 91} (2015) 024017}
  [\href{https://arxiv.org/abs/1410.5348}{{\ttfamily 1410.5348}}].

\bibitem{BjerrumBohr:2002ks}
N.~E.~J. Bjerrum-Bohr, J.~F. Donoghue and B.~R. Holstein, \emph{{Quantum
  corrections to the Schwarzschild and Kerr metrics}},
  \href{https://doi.org/10.1103/PhysRevD.68.084005}{\emph{Phys. Rev. D}
  {\bfseries 68} (2003) 084005}
  [\href{https://arxiv.org/abs/hep-th/0211071}{{\ttfamily hep-th/0211071}}].

\bibitem{LIGOGW}
{B. P. Abbott et al., LIGO Scientific Collaboration and Virgo Collaboration},
  \emph{Observation of gravitational waves from a binary black hole merger},
  \href{https://doi.org/10.1103/PhysRevLett.116.061102}{\emph{Phys. Rev. Lett.}
  {\bfseries 116} (2016) 061102}.

\bibitem{BERTOTTI1956}
B.~Bertotti, \emph{On gravitational motion},
  \href{https://doi.org/https://doi.org/10.1007/BF02746175}{\emph{Il Nuovo
  Cimento} (1956) }.

\bibitem{BERTOTTI1960169}
B.~Bertotti and J.~Plebanski, \emph{Theory of gravitational perturbations in
  the fast motion approximation},
  \href{https://doi.org/https://doi.org/10.1016/0003-4916(60)90132-9}{\emph{Annals
  of Physics} {\bfseries 11} (1960) 169 }.

\bibitem{Cheung:2018wkq}
C.~Cheung, I.~Z. Rothstein and M.~P. Solon, \emph{{From Scattering Amplitudes
  to Classical Potentials in the Post-Minkowskian Expansion}},
  \href{https://doi.org/10.1103/PhysRevLett.121.251101}{\emph{Phys. Rev. Lett.}
  {\bfseries 121} (2018) 251101}
  [\href{https://arxiv.org/abs/1808.02489}{{\ttfamily 1808.02489}}].

\bibitem{Cristofoli:2019neg}
A.~Cristofoli, N.~E.~J. Bjerrum-Bohr, P.~H. Damgaard and P.~Vanhove, \emph{{On
  Post-Minkowskian Hamiltonians in General Relativity}},
  \href{https://arxiv.org/abs/1906.01579}{{\ttfamily 1906.01579}}.

\bibitem{Bern:2020buy}
Z.~Bern, A.~Luna, R.~Roiban, C.-H. Shen and M.~Zeng, \emph{{Spinning Black Hole
  Binary Dynamics, Scattering Amplitudes and Effective Field Theory}},
  \href{https://arxiv.org/abs/2005.03071}{{\ttfamily 2005.03071}}.

\bibitem{Kosower:2018adc}
D.~A. Kosower, B.~Maybee and D.~O'Connell, \emph{{Amplitudes, Observables, and
  Classical Scattering}},
  \href{https://doi.org/10.1007/JHEP02(2019)137}{\emph{JHEP} {\bfseries 02}
  (2019) 137} [\href{https://arxiv.org/abs/1811.10950}{{\ttfamily
  1811.10950}}].

\bibitem{Maybee:2019jus}
B.~Maybee, D.~O'Connell and J.~Vines, \emph{{Observables and amplitudes for
  spinning particles and black holes}},
  \href{https://doi.org/10.1007/JHEP12(2019)156}{\emph{JHEP} {\bfseries 12}
  (2019) 156} [\href{https://arxiv.org/abs/1906.09260}{{\ttfamily
  1906.09260}}].

\bibitem{Kalin:2019rwq}
G.~Kälin and R.~A. Porto, \emph{{From Boundary Data to Bound States}},
  \href{https://arxiv.org/abs/1910.03008}{{\ttfamily 1910.03008}}.

\bibitem{Bjerrum-Bohr:2019kec}
N.~E.~J. Bjerrum-Bohr, A.~Cristofoli and P.~H. Damgaard,
  \emph{{Post-Minkowskian Scattering Angle in Einstein Gravity}},
  \href{https://arxiv.org/abs/1910.09366}{{\ttfamily 1910.09366}}.

\bibitem{Cristofoli:2020uzm}
A.~Cristofoli, P.~H. Damgaard, P.~Di~Vecchia and C.~Heissenberg,
  \emph{{Second-order Post-Minkowskian scattering in arbitrary dimensions}},
  \href{https://doi.org/10.1007/JHEP07(2020)122}{\emph{JHEP} {\bfseries 07}
  (2020) 122} [\href{https://arxiv.org/abs/2003.10274}{{\ttfamily
  2003.10274}}].

\bibitem{Cristofoli:2020hnk}
A.~Cristofoli, \emph{{Gravitational shock waves and scattering amplitudes}},
  \href{https://arxiv.org/abs/2006.08283}{{\ttfamily 2006.08283}}.

\bibitem{Zvi3PM}
Z.~Bern, C.~Cheung, R.~Roiban, C.-H. Shen, M.~P. Solon and M.~Zeng,
  \emph{{Scattering Amplitudes and the Conservative Hamiltonian for Binary
  Systems at Third Post-Minkowskian Order}},
  \href{https://arxiv.org/abs/1901.04424}{{\ttfamily 1901.04424}}.

\bibitem{Bern:2019crd}
Z.~Bern, C.~Cheung, R.~Roiban, C.-H. Shen, M.~P. Solon and M.~Zeng,
  \emph{{Black Hole Binary Dynamics from the Double Copy and Effective
  Theory}},  \href{https://arxiv.org/abs/1908.01493}{{\ttfamily 1908.01493}}.

\bibitem{Cheung:2020gyp}
C.~Cheung and M.~P. Solon, \emph{{Classical gravitational scattering at $
  \mathcal{O} $(G$^{3}$) from Feynman diagrams}},
  \href{https://doi.org/10.1007/JHEP06(2020)144}{\emph{JHEP} {\bfseries 06}
  (2020) 144} [\href{https://arxiv.org/abs/2003.08351}{{\ttfamily
  2003.08351}}].

\bibitem{Cheung:2020sdj}
C.~Cheung and M.~P. Solon, \emph{{Tidal Effects in the Post-Minkowskian
  Expansion}},  \href{https://arxiv.org/abs/2006.06665}{{\ttfamily
  2006.06665}}.

\bibitem{Bern:2020gjj}
Z.~Bern, H.~Ita, J.~Parra-Martinez and M.~S. Ruf, \emph{{Universality in the
  classical limit of massless gravitational scattering}},
  \href{https://doi.org/10.1103/PhysRevLett.125.031601}{\emph{Phys. Rev. Lett.}
  {\bfseries 125} (2020) 031601}
  [\href{https://arxiv.org/abs/2002.02459}{{\ttfamily 2002.02459}}].

\bibitem{Brandhuber:2019qpg}
A.~Brandhuber and G.~Travaglini, \emph{{On higher-derivative effects on the
  gravitational potential and particle bending}},
  \href{https://doi.org/10.1007/JHEP01(2020)010}{\emph{JHEP} {\bfseries 01}
  (2020) 010} [\href{https://arxiv.org/abs/1905.05657}{{\ttfamily
  1905.05657}}].

\bibitem{Cristofoli:2019ewu}
A.~Cristofoli, \emph{{Post-Minkowskian Hamiltonians in Modified Theories of
  Gravity}}, \href{https://doi.org/10.1016/j.physletb.2019.135095}{\emph{Phys.
  Lett. B} {\bfseries 800} (2020) 135095}
  [\href{https://arxiv.org/abs/1906.05209}{{\ttfamily 1906.05209}}].

\bibitem{Damgaard:2019lfh}
P.~H. Damgaard, K.~Haddad and A.~Helset, \emph{{Heavy Black Hole Effective
  Theory}}, \href{https://doi.org/10.1007/JHEP11(2019)070}{\emph{JHEP}
  {\bfseries 11} (2019) 070}
  [\href{https://arxiv.org/abs/1908.10308}{{\ttfamily 1908.10308}}].

\bibitem{Guevara:2017csg}
A.~Guevara, \emph{{Holomorphic Classical Limit for Spin Effects in
  Gravitational and Electromagnetic Scattering}},
  \href{https://doi.org/10.1007/JHEP04(2019)033}{\emph{JHEP} {\bfseries 04}
  (2019) 033} [\href{https://arxiv.org/abs/1706.02314}{{\ttfamily
  1706.02314}}].

\bibitem{Arkani-Hamed:2017jhn}
N.~Arkani-Hamed, T.-C. Huang and Y.-t. Huang, \emph{{Scattering Amplitudes For
  All Masses and Spins}},  \href{https://arxiv.org/abs/1709.04891}{{\ttfamily
  1709.04891}}.

\bibitem{Guevara:2018wpp}
A.~Guevara, A.~Ochirov and J.~Vines, \emph{{Scattering of Spinning Black Holes
  from Exponentiated Soft Factors}},
  \href{https://doi.org/10.1007/JHEP09(2019)056}{\emph{JHEP} {\bfseries 09}
  (2019) 056} [\href{https://arxiv.org/abs/1812.06895}{{\ttfamily
  1812.06895}}].

\bibitem{Chung:2018kqs}
M.-Z. Chung, Y.-T. Huang, J.-W. Kim and S.~Lee, \emph{{The simplest massive
  S-matrix: from minimal coupling to Black Holes}},
  \href{https://doi.org/10.1007/JHEP04(2019)156}{\emph{JHEP} {\bfseries 04}
  (2019) 156} [\href{https://arxiv.org/abs/1812.08752}{{\ttfamily
  1812.08752}}].

\bibitem{Chung:2019duq}
M.-Z. Chung, Y.-T. Huang and J.-W. Kim, \emph{{Classical potential for general
  spinning bodies}},  \href{https://arxiv.org/abs/1908.08463}{{\ttfamily
  1908.08463}}.

\bibitem{Guevara:2019fsj}
A.~Guevara, A.~Ochirov and J.~Vines, \emph{{Black-hole scattering with general
  spin directions from minimal-coupling amplitudes}},
  \href{https://arxiv.org/abs/1906.10071}{{\ttfamily 1906.10071}}.

\bibitem{Arkani-Hamed:2019ymq}
N.~Arkani-Hamed, Y.-t. Huang and D.~O'Connell, \emph{{Kerr Black Holes as
  Elementary Particles}},  \href{https://arxiv.org/abs/1906.10100}{{\ttfamily
  1906.10100}}.

\bibitem{Aoude:2020onz}
R.~Aoude, K.~Haddad and A.~Helset, \emph{{On-shell heavy particle effective
  theories}}, \href{https://doi.org/10.1007/JHEP05(2020)051}{\emph{JHEP}
  {\bfseries 05} (2020) 051}
  [\href{https://arxiv.org/abs/2001.09164}{{\ttfamily 2001.09164}}].

\bibitem{Chung:2020rrz}
M.-Z. Chung, Y.-t. Huang, J.-W. Kim and S.~Lee, \emph{{Complete Hamiltonian for
  spinning binary systems at first post-Minkowskian order}},
  \href{https://doi.org/10.1007/JHEP05(2020)105}{\emph{JHEP} {\bfseries 05}
  (2020) 105} [\href{https://arxiv.org/abs/2003.06600}{{\ttfamily
  2003.06600}}].

\bibitem{Aoude:2020mlg}
R.~Aoude, M.-Z. Chung, Y.-t. Huang, C.~S. Machado and M.-K. Tam, \emph{{The
  silence of binary Kerr}},  \href{https://arxiv.org/abs/2007.09486}{{\ttfamily
  2007.09486}}.

\bibitem{Damour:1992qi}
T.~Damour, M.~Soffel and C.-m. Xu, \emph{{General relativistic celestial
  mechanics. 3. Rotational equations of motion}},
  \href{https://doi.org/10.1103/PhysRevD.47.3124}{\emph{Phys. Rev. D}
  {\bfseries 47} (1993) 3124}.

\bibitem{Damour:2009wj}
T.~Damour and A.~Nagar, \emph{{Effective One Body description of tidal effects
  in inspiralling compact binaries}},
  \href{https://doi.org/10.1103/PhysRevD.81.084016}{\emph{Phys. Rev. D}
  {\bfseries 81} (2010) 084016}
  [\href{https://arxiv.org/abs/0911.5041}{{\ttfamily 0911.5041}}].

\bibitem{Buonanno:1998gg}
A.~Buonanno and T.~Damour, \emph{{Effective one-body approach to general
  relativistic two-body dynamics}},
  \href{https://doi.org/10.1103/PhysRevD.59.084006}{\emph{Phys. Rev. D}
  {\bfseries 59} (1999) 084006}
  [\href{https://arxiv.org/abs/gr-qc/9811091}{{\ttfamily gr-qc/9811091}}].

\bibitem{Bini:2012gu}
D.~Bini, T.~Damour and G.~Faye, \emph{{Effective action approach to
  higher-order relativistic tidal interactions in binary systems and their
  effective one body description}},
  \href{https://doi.org/10.1103/PhysRevD.85.124034}{\emph{Phys. Rev. D}
  {\bfseries 85} (2012) 124034}
  [\href{https://arxiv.org/abs/1202.3565}{{\ttfamily 1202.3565}}].

\bibitem{Bini:2020flp}
D.~Bini, T.~Damour and A.~Geralico, \emph{{Scattering of tidally interacting
  bodies in post-Minkowskian gravity}},
  \href{https://doi.org/10.1103/PhysRevD.101.044039}{\emph{Phys. Rev. D}
  {\bfseries 101} (2020) 044039}
  [\href{https://arxiv.org/abs/2001.00352}{{\ttfamily 2001.00352}}].

\bibitem{Kalin:2020mvi}
G.~Kälin and R.~A. Porto, \emph{{Post-Minkowskian Effective Field Theory for
  Conservative Binary Dynamics}},
  \href{https://arxiv.org/abs/2006.01184}{{\ttfamily 2006.01184}}.

\bibitem{Henry:2019xhg}
Q.~Henry, G.~Faye and L.~Blanchet, \emph{{Tidal effects in the equations of
  motion of compact binary systems to next-to-next-to-leading post-Newtonian
  order}}, \href{https://doi.org/10.1103/PhysRevD.101.064047}{\emph{Phys. Rev.
  D} {\bfseries 101} (2020) 064047}
  [\href{https://arxiv.org/abs/1912.01920}{{\ttfamily 1912.01920}}].

\bibitem{Benvenuti:2006qr}
S.~Benvenuti, B.~Feng, A.~Hanany and Y.-H. He, \emph{{Counting BPS Operators in
  Gauge Theories: Quivers, Syzygies and Plethystics}},
  \href{https://doi.org/10.1088/1126-6708/2007/11/050}{\emph{JHEP} {\bfseries
  11} (2007) 050} [\href{https://arxiv.org/abs/hep-th/0608050}{{\ttfamily
  hep-th/0608050}}].

\bibitem{Feng:2007ur}
B.~Feng, A.~Hanany and Y.-H. He, \emph{{Counting gauge invariants: The
  Plethystic program}},
  \href{https://doi.org/10.1088/1126-6708/2007/03/090}{\emph{JHEP} {\bfseries
  03} (2007) 090} [\href{https://arxiv.org/abs/hep-th/0701063}{{\ttfamily
  hep-th/0701063}}].

\bibitem{Jenkins:2009dy}
E.~E. Jenkins and A.~V. Manohar, \emph{{Algebraic Structure of Lepton and Quark
  Flavor Invariants and CP Violation}},
  \href{https://doi.org/10.1088/1126-6708/2009/10/094}{\emph{JHEP} {\bfseries
  10} (2009) 094} [\href{https://arxiv.org/abs/0907.4763}{{\ttfamily
  0907.4763}}].

\bibitem{Lehman:2015via}
L.~Lehman and A.~Martin, \emph{{Hilbert Series for Constructing Lagrangians:
  expanding the phenomenologist's toolbox}},
  \href{https://doi.org/10.1103/PhysRevD.91.105014}{\emph{Phys. Rev. D}
  {\bfseries 91} (2015) 105014}
  [\href{https://arxiv.org/abs/1503.07537}{{\ttfamily 1503.07537}}].

\bibitem{Lehman:2015coa}
L.~Lehman and A.~Martin, \emph{{Low-derivative operators of the Standard Model
  effective field theory via Hilbert series methods}},
  \href{https://doi.org/10.1007/JHEP02(2016)081}{\emph{JHEP} {\bfseries 02}
  (2016) 081} [\href{https://arxiv.org/abs/1510.00372}{{\ttfamily
  1510.00372}}].

\bibitem{Henning:2015alf}
B.~Henning, X.~Lu, T.~Melia and H.~Murayama, \emph{{2, 84, 30, 993, 560, 15456,
  11962, 261485, ...: Higher dimension operators in the SM EFT}},
  \href{https://doi.org/10.1007/JHEP08(2017)016}{\emph{JHEP} {\bfseries 08}
  (2017) 016} [\href{https://arxiv.org/abs/1512.03433}{{\ttfamily
  1512.03433}}].

\bibitem{Henning:2015daa}
B.~Henning, X.~Lu, T.~Melia and H.~Murayama, \emph{{Hilbert series and operator
  bases with derivatives in effective field theories}},
  \href{https://doi.org/10.1007/s00220-015-2518-2}{\emph{Commun. Math. Phys.}
  {\bfseries 347} (2016) 363}
  [\href{https://arxiv.org/abs/1507.07240}{{\ttfamily 1507.07240}}].

\bibitem{Henning:2017fpj}
B.~Henning, X.~Lu, T.~Melia and H.~Murayama, \emph{{Operator bases,
  $S$-matrices, and their partition functions}},
  \href{https://doi.org/10.1007/JHEP10(2017)199}{\emph{JHEP} {\bfseries 10}
  (2017) 199} [\href{https://arxiv.org/abs/1706.08520}{{\ttfamily
  1706.08520}}].

\bibitem{Ruhdorfer:2019qmk}
M.~Ruhdorfer, J.~Serra and A.~Weiler, \emph{{Effective Field Theory of Gravity
  to All Orders}}, \href{https://doi.org/10.1007/JHEP05(2020)083}{\emph{JHEP}
  {\bfseries 05} (2020) 083}
  [\href{https://arxiv.org/abs/1908.08050}{{\ttfamily 1908.08050}}].

\bibitem{Bern:2020uwk}
Z.~Bern, J.~Parra-Martinez, R.~Roiban, E.~Sawyer and C.-H. Shen, \emph{{Leading
  Nonlinear Tidal Effects and Scattering Amplitudes}},
  \href{https://arxiv.org/abs/2010.08559}{{\ttfamily 2010.08559}}.

\bibitem{Poplawski:2009fb}
N.~J. Poplawski, \emph{{Classical Physics: Spacetime and Fields}},
  \href{https://arxiv.org/abs/0911.0334}{{\ttfamily 0911.0334}}.

\bibitem{Guth:2014hsa}
A.~H. Guth, M.~P. Hertzberg and C.~Prescod-Weinstein, \emph{{Do Dark Matter
  Axions Form a Condensate with Long-Range Correlation?}},
  \href{https://doi.org/10.1103/PhysRevD.92.103513}{\emph{Phys. Rev. D}
  {\bfseries 92} (2015) 103513}
  [\href{https://arxiv.org/abs/1412.5930}{{\ttfamily 1412.5930}}].

\bibitem{Namjoo:2017nia}
M.~H. Namjoo, A.~H. Guth and D.~I. Kaiser, \emph{{Relativistic Corrections to
  Nonrelativistic Effective Field Theories}},
  \href{https://doi.org/10.1103/PhysRevD.98.016011}{\emph{Phys. Rev. D}
  {\bfseries 98} (2018) 016011}
  [\href{https://arxiv.org/abs/1712.00445}{{\ttfamily 1712.00445}}].

\bibitem{Braaten:2018lmj}
E.~Braaten, A.~Mohapatra and H.~Zhang, \emph{{Classical Nonrelativistic
  Effective Field Theories for a Real Scalar Field}},
  \href{https://doi.org/10.1103/PhysRevD.98.096012}{\emph{Phys. Rev. D}
  {\bfseries 98} (2018) 096012}
  [\href{https://arxiv.org/abs/1806.01898}{{\ttfamily 1806.01898}}].

\bibitem{Haddad:2020tvs}
K.~Haddad and A.~Helset, \emph{{The double copy for heavy particles}},
  \href{https://arxiv.org/abs/2005.13897}{{\ttfamily 2005.13897}}.

\bibitem{PASSARINO1979151}
G.~Passarino and M.~Veltman, \emph{One-loop corrections for $e+e-$ annihilation
  into $\mu+\mu-$ in the weinberg model},
  \href{https://doi.org/https://doi.org/10.1016/0550-3213(79)90234-7}{\emph{Nuclear
  Physics B} {\bfseries 160} (1979) 151 }.

\bibitem{Goldberger:2004jt}
W.~D. Goldberger and I.~Z. Rothstein, \emph{{An Effective field theory of
  gravity for extended objects}},
  \href{https://doi.org/10.1103/PhysRevD.73.104029}{\emph{Phys. Rev. D}
  {\bfseries 73} (2006) 104029}
  [\href{https://arxiv.org/abs/hep-th/0409156}{{\ttfamily hep-th/0409156}}].

\bibitem{Barabanschikov:2005ri}
A.~Barabanschikov, L.~Grant, L.~L. Huang and S.~Raju, \emph{{The Spectrum of
  Yang Mills on a sphere}},
  \href{https://doi.org/10.1088/1126-6708/2006/01/160}{\emph{JHEP} {\bfseries
  01} (2006) 160} [\href{https://arxiv.org/abs/hep-th/0501063}{{\ttfamily
  hep-th/0501063}}].

\bibitem{Huber:2019ugz}
M.~Accettulli~Huber, A.~Brandhuber, S.~De~Angelis and G.~Travaglini,
  \emph{{Note on the absence of $R^2$ corrections to Newton's potential}},
  \href{https://doi.org/10.1103/PhysRevD.101.046011}{\emph{Phys. Rev. D}
  {\bfseries 101} (2020) 046011}
  [\href{https://arxiv.org/abs/1911.10108}{{\ttfamily 1911.10108}}].

\bibitem{Hays:2018zze}
C.~Hays, A.~Martin, V.~Sanz and J.~Setford, \emph{{On the impact of
  dimension-eight SMEFT operators on Higgs measurements}},
  \href{https://doi.org/10.1007/JHEP02(2019)123}{\emph{JHEP} {\bfseries 02}
  (2019) 123} [\href{https://arxiv.org/abs/1808.00442}{{\ttfamily
  1808.00442}}].

\bibitem{Shadmi:2018xan}
Y.~Shadmi and Y.~Weiss, \emph{{Effective Field Theory Amplitudes the On-Shell
  Way: Scalar and Vector Couplings to Gluons}},
  \href{https://doi.org/10.1007/JHEP02(2019)165}{\emph{JHEP} {\bfseries 02}
  (2019) 165} [\href{https://arxiv.org/abs/1809.09644}{{\ttfamily
  1809.09644}}].

\end{thebibliography}\endgroup

\end{document}